\documentclass[12pt]{JHEP3}

\usepackage{amscd,amsmath,amssymb,amsfonts,xspace,mathrsfs,amsthm}
\usepackage{color, mathtools}
\usepackage[dvipsnames]{xcolor}

\usepackage{url}
\let\normalcolor\relax

\usepackage{latexsym}
\usepackage{graphicx}
\usepackage{dsfont}
\usepackage{longtable}
\usepackage{enumitem}

\usepackage[latin1]{inputenc}
\usepackage{multirow,float}

\usepackage{fancybox}
\usepackage{bbm}
\usepackage{bm}
\usepackage{booktabs}
\usepackage{soul}
\usepackage{physics}

\usepackage[numbers,compress]{natbib}

\usepackage{tikz}
\usetikzlibrary{decorations.pathmorphing,arrows.meta}

\tikzset{
  fermion/.style={-Latex, thick},
  gauge/.style={decorate, decoration={snake}, thick},
  tensor/.style={dashed, thick},
}

\usepackage{blkarray}

\usepackage{empheq}

\numberwithin{equation}{section}

\def\bea{\begin{eqnarray}}
\def\eea{\end{eqnarray}}
\def\be{\begin{equation}}
\def\ee{\end{equation}}
\def\ba{\begin{align}}
\def\ea{\end{align}}
\def\bse{\begin{subequations}}
\def\ese{\end{subequations}}

\def\({\left(}
\def\){\right)}
\def\[{\left[}
\def\]{\right]}
\def\<{\left\langle}
\def\>{\right\rangle}

\newcommand{\binomial}[2]{ \left( {#1 \atop #2} \right)}

\def\sf#1{{\color{green} #1}}

\def\BM{\begin{matrix}}
\def\EM{\end{matrix}}

\title{
Supergravity anomaly equations from modularity of Calabi--Yau threefolds}

\author{Cesar Fierro Cota$^1$,
\\
$^1${\it 
Laboratoire de Physique Th\'eorique et Hautes Energies, 
CNRS and Sorbonne Universit\'e, 
Campus Pierre et Marie Curie, 4 place Jussieu, F-75005 Paris, France}

\vspace*{2mm} {\tt e-mail:
\email{fierrocota@lpthe.jussieu.fr}
}

\vspace*{-3mm}

}

\abstract{F-theory compactifications on elliptically fibered Calabi--Yau threefolds yield consistent six-dimensional $\mathcal{N}=(1,0)$ supergravity theories, for which the cancellation of gravitational, gauge and mixed anomalies imposes non-trivial algebraic relations between classical intersection data and enumerative geometry invariants of curves in the fiber. In this work, we capture the spectrum of such theories via meromorphic quasi-Jacobi forms of index zero whose Fourier coefficients determine the genus zero Gromov--Witten theory restricted to curve classes in the fiber. We find that the one-loop anomaly coefficients of the effective six-dimensional theories are encoded in the modular properties of these automorphic forms, while the Green--Schwarz counterterms are made manifest by the Fourier--Mukai transform action on zero- and two-branes associated with double T-duality along the elliptic fiber.
Moreover, we show that the anomaly cancellation conditions are automatically satisfied in this class of string compactifications as a consequence of the holomorphic anomaly equations of topological string theory.}

\begin{document}
\setlength{\parskip}{0.2cm}

\section{Introduction}

One of the major achievements of string theory is the realization of supergravity and quantum field theories that are free of quantum anomalies~\cite{Green:1984sg}. In six-dimensional supergravity theories with minimal supersymmetry, the absence of one-loop perturbative  anomalies can be achieved  by adding appropriate Green--Schwarz local counterterms~\cite{Green:1984bx,Sagnotti:1992qw,Sadov:1996zm}. 
The resulting anomaly cancellation conditions translate into strong algebraic constraints that relate the massless spectrum of the theory to the so-called gauge and gravitational anomaly coefficients, which are real-valued vectors transforming under $SO(1,n_T)$, where $n_T$ denotes the number of tensor multiplets. 

F-theory is a non-perturbative formulation of Type IIB string theory that geometrizes backgrounds with backreacting 7-branes~\cite{Vafa:1996xn,Weigand:2018rez,Cvetic:2018bni}. It incorporates the presence of a varying axio-dilaton profile over a  complex K\"ahler manifold $B$, 
which is complex-valued and defined only up to $\mathrm{S}$-duality transformations in $\mathrm{SL}(2,\mathbb{Z})$. Accordingly, one identifies the axio-dilaton with the complex structure modulus of an elliptic curve, giving rise to the structure of an elliptically fibered Calabi--Yau manifold $\pi: X \to B$. This setup, 
Type IIB on $B$, yields an effective field theory whose relevant data 
are encapsulated by algebraic-geometric properties of $X$, including in particular the realization of anomalies via the Green--Schwarz cancellation mechanism.

We are interested in discussing F-theory on smooth, compact, elliptically fibered Calabi--Yau threefolds $\pi: X\to B$, which engineer six-dimensional supergravity theories with $\mathcal{N}=(1,0)$ supersymmetry.  In this setting, the Green--Schwarz anomaly cancellation conditions relate two types of data:
\begin{enumerate}
    \item The anomaly coefficients $\{b\}$, 
    determined by the intersection data of $X$~\cite{Park:2011ji}. 
    \item The massless spectrum $S$,  determined by the enumerative geometry of $X$~\cite{Paul-KonstantinOehlmann:2019jgr,Kashani-Poor:2019jyo}.
\end{enumerate}
The anomaly cancellation conditions can be written schematically as
\begin{equation}
    \mathcal{A}_i^{\mathrm{GS}}(\{b\}) = \mathcal{A}^{\mathrm{loop}}_i(S)\,,
    \label{eqn:Anomalies}
\end{equation}
where $\mathcal{A}_i^{\mathrm{loop}}$ and $\mathcal{A}^{\mathrm{GS}}_i$ are some functions of the geometric data. 
Consequently, the physics-geometry dictionary implies non-trivial identities among the intersection numbers and enumerative invariants of $X$. 

The purpose of this work is to explain these physical predictions from geometry and to elucidate the geometrical realization of the Green--Schwarz mechanism. Our central claim is that modularity captures these physics constraints through recursive relations that tie quantum geometry data---namely Gromov--Witten invariants---to the classical intersection theory of a given elliptically fibered Calabi--Yau threefold. 

Already in the early days of string theory, it was observed that one-loop modular invariance in perturbative Type IIB and heterotic string theory ensures the validity of the Green--Schwarz mechanism~\cite{Schellekens:1986yi,Schellekens:1986xh}.  
Moreover, it was understood that the elliptic genera of heterotic strings in even dimensions encode the one-loop anomalies and the corresponding Green-Schwarz counterterms of the associated effective field theories~\cite{Lerche:1987qk}. 
More recently, in F-theory compactifications admitting a heterotic dual description, the elliptic genera of heterotic strings have been identified with the reduced Gromov--Witten theory on special K3 fiber divisors in Calabi--Yau manifolds that admit a K3 fibration~\cite{Klemm:1996hh,Lee:2018urn,Lee:2018spm,Lee:2019tst,Lee:2020gvu,Lee:2020blx,Maulik:2007mp}. In these non-perturbative configurations, explicit examples show that the Green-Schwarz mechanism is likewise derived from the elliptic genera of heterotic strings in six- and four-dimensions~\cite{Lee:2020gvu}, described in terms of meromorphic Jacobi and quasi-Jacobi forms~\cite{Lee:2020gvu,Lee:2020blx}, respectively. 

Jacobi forms are holomorphic functions of two complex variables that generalize modular forms and satisfy modular and elliptic transformation laws analogous to those of Jacobi theta functions~\cite{eichler1985theory}. In contrast, quasimodular forms are holomorphic functions on the complex upper half-plane that extend the notion of modular forms by satisfying modified  ``anomalous" transformation properties under $\mathrm{SL}(2,\mathbb{Z})$~\cite{10.1007/978-1-4612-4264-2_6}. 
A more general class of automorphic objects is that of quasi-Jacobi forms---capturing both Jacobi and quasimodular forms---which are also holomorphic functions of two complex variables that satisfy both modular and elliptic transformation properties up to anomalous corrections~\cite{Libgober2009EllipticGR,Oberdieck:2017pqm,Oberdieck:2022khj}. 

For torus fibered Calabi--Yau threefolds, it is conjectured that the topological string partition function can be expanded in a Fourier series over the base K\"ahler moduli, with coefficients given by  meromorphic Jacobi forms~\cite{Huang:2015ada,Gu:2017ccq,DelZotto:2017mee,Cota:2019cjx}. See~\cite{Pioline:2025uov} for recent progress towards a proof of the original conjecture of~\cite{Huang:2015ada}. A similar structure appears for elliptically fibered Calabi--Yau fourfolds,  in which the corresponding BPS generating functions admit an expansion whose coefficients are meromorphic quasi-Jacobi forms~\cite{Haghighat:2015qdq,Cota:2017aal,Lee:2019tst,Lee:2020gvu,Lee:2020blx}. In this work, quasi-Jacobi forms reappear in the context of mirror symmetry for Calabi--Yau threefolds, where they embody the automorphic forms underlying the topological $n$-point correlation functions---in accordance with conjectures of~\cite{Oberdieck:2017pqm}. As we will see, it is precisely the anomalous modular transformation behavior of quasi-Jacobi forms that provides the mechanism from which the anomaly cancellation conditions~(\ref{eqn:Anomalies}) follow. We summarize this result next.

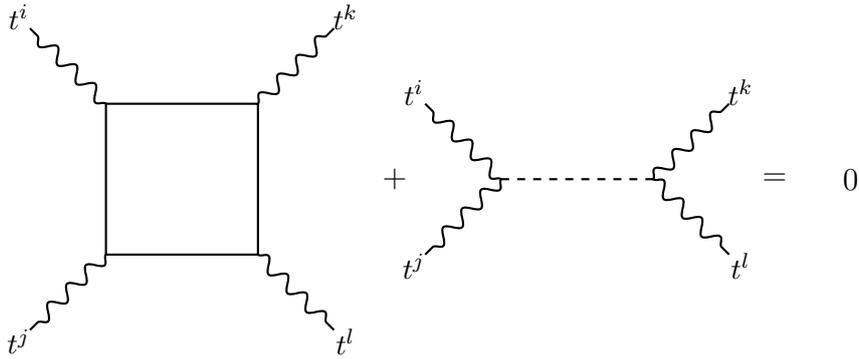
\begin{figure}[t]
\centering
\begin{tikzpicture}[x=1cm,y=1cm,baseline=(current bounding box.center)]
  \usetikzlibrary{decorations.pathmorphing,arrows.meta}

  \tikzset{
    fermion/.style={-Latex, thick},
    gauge/.style={decorate, decoration={snake}, thick},
    tensor/.style={dashed, thick},
  }

  \begin{scope}[shift={(0,0)}]
    \coordinate (v1) at (-1,-1);
    \coordinate (v2) at ( 1,-1);
    \coordinate (v3) at ( 1, 1);
    \coordinate (v4) at (-1, 1);

    \draw[fermion] (v1) -- (v2) -- (v3) -- (v4) -- cycle;

    \coordinate (e1) at (-2,  2); 
    \coordinate (e2) at ( 2,  2); 
    \coordinate (e3) at ( 2, -2); 
    \coordinate (e4) at (-2, -2); 

    \draw[gauge] (v4) -- (e1) node[pos=1.15, font=\small] {$t^i$};
    \draw[gauge] (v3) -- (e2) node[pos=1.15, font=\small] {$t^k$};
    \draw[gauge] (v2) -- (e3) node[pos=1.15, font=\small] {$t^l$};
    \draw[gauge] (v1) -- (e4) node[pos=1.15, font=\small] {$t^j$};
  \end{scope}

  \node at (2.8,0) {$+$};

  \begin{scope}[shift={(5.2,0)}]
    \coordinate (vL) at (-1,0);
    \coordinate (vR) at ( 1,0);

    \coordinate (LT) at (-2, 1);
    \coordinate (LB) at (-2,-1);
    \draw[gauge] (vL) -- (LT) node[pos=1.15, font=\small] {$t^i$};
    \draw[gauge] (vL) -- (LB) node[pos=1.15, font=\small] {$t^j$};

    \coordinate (RT) at ( 2, 1);
    \coordinate (RB) at ( 2,-1);
    \draw[gauge] (vR) -- (RT) node[pos=1.15, font=\small] {$t^k$};
    \draw[gauge] (vR) -- (RB) node[pos=1.15, font=\small] {$t^l$};

    \draw[tensor] (vL) -- (vR)
      node[midway, above=2pt, font=\small] {};
  \end{scope}

  \node at (7.8,0) {$=$};
  \node at (8.8,0) {$0$};

\end{tikzpicture}
 \caption{Schematic representation of Green--Schwarz mechanism and  equation~(\ref{eqn:modAnomlay}). We associate $\tau$ with external graviton insertions and $z^A$ with external gauge boson insertions. The topological four-point couplings $C_{ABCD}$ correspond to the pure gauge anomalies, $C_{\tau\tau AB}$ to the mixed anomalies and $C_{\tau\tau\tau\tau}$ to the pure gravitational anomaly.}
 \label{Fig:1}
\end{figure}

Let us consider a smooth elliptically fibered Calabi--Yau threefold $\pi: X \to B$ obtained as a resolution of an elliptic fibration with Kodaira singular fibers carrying an associated Lie algebra $\mathfrak{g}$.
Denote by $\tau$ the complexified K\"ahler volume of the elliptic fiber and by  $z^{A}$ the complexified K\"ahler volume of isolated fibral curves $C^{A}\subset X$ appearing over points in $B$ corresponding to codimension-two singularities enhancements, where $A = 1, \ldots , \mathrm{rk}(\mathfrak{g})$. These K\"ahler moduli parametrise the K\"ahler form  $\omega = t^i \omega_i=\tau \omega_0+z^A \omega_A+t^\alpha \omega_\alpha$ of $X$ along the fiber directions, where $\{\omega_i\}$ is a basis of $H^{1,1}(X)$, $t^i$ are the corresponding complexified  K\"ahler moduli and  $\alpha = 1, \ldots , h^{1,1}(B)$. 
In the A-model topological string theory, the topological  $n$-point correlation functions $C_{i_1 \cdots i_n}$ require  $n$ insertions of chiral operators associated with the classes  $\omega_{i}$. 
The anomaly cancellation conditions~(\ref{eqn:Anomalies}) appear in the zero-th Fourier coefficient of a modular anomaly equation of the form
\begin{equation}
\frac{\partial}{\partial E_2}C_{ijkl} + C^{\mathrm{GS}}_{ijkl} =0\,,
\label{eqn:modAnomlay}
\end{equation}
where we take purely  insertions for the four-point coupling $C_{ijkl}$ of the type $t^i \in\{\tau,z^{A}\}$, which behave as a quasi-Jacobi forms of weight $k$ and index zero. 
Here $E_2$ is the second Eisenstein series, a quasi-modular form~\cite{10.1007/978-1-4612-4264-2_6}, and $C_{ijkl}^{\mathrm{GS}}$ is a four-tensor encoding the Green--Schwarz classical counterterms that 
behaves as a quasi-Jacobi form of weight $k-2$ and index zero. The modular anomaly operator $\partial_{E_2}$ measures the one-loop anomaly encoded in the four-point function $C_{ijkl}$ as we depict in Figure~\ref{Fig:1}.
As we will see, equation~(\ref{eqn:modAnomlay}) is a consequence of the holomorphic anomaly equations by~\cite{Bershadsky:1993cx}.


 We now outline and explain the roadmap of this work. In section~\ref{sec:2} we provide a brief review for $\mathcal{N}=(1,0)$ six-dimensional supergravity theories and the necessary physics-geometry dictionary of elliptic fibrations for the subsequent chapters. 
In section~\ref{sec:3} we introduce some elements of mirror and homological symmetry to argue for the modularity associated to the enumerative geometry of elliptically fibered Calabi--Yau threefolds. 
The main part of this work is section~\ref{sec:4}. We provide examples in section~\ref{sec:eg}. Finally, we provide our conclusions and further discussion in section~\ref{sec:6}. Moreover, we include a brief review on Gromov--Witten theory in Appendix~\ref{App:GW}. In Appendix~\ref{app:GSanomalies} we add complementary data. We also review the necessary material for quasi-Jacobi forms in Appendix~\ref{app:QJac}. In Appendix~\ref{App:WeylChar}, we introduce Weyl characters.
\\\\
\noindent\textbf{Acknowledgements:} We thank Boris Pioline, Thorsten Schimannek, Amir-Kian Kashani-Poor, Albrecht Klemm and Timo Weigand for valuable discussions and for inspiring this work. The research of CFC is supported by the Initiative Physique des Infinis at Sorbonne Universit\'e.

\section{Six-dimensional supergravities}
\label{sec:2}

In this section, we briefly review the Green--Schwarz mechanism and the anomaly cancellation conditions, followed by a review of F-theory in which we define key geometric objects of elliptic fibrations that will be necessary for the subsequent chapters. 

\subsection{Review of anomaly equations}
\label{sec:Anomalies}

   As a starting point, we summarize the massless spectrum content in a six-dimensional  supergravity theory. This is given by the following $\mathcal{N} = (1,0)$ supermultiplets:~\cite{Taylor:2011wt}
   \begin{itemize}
           \item \textbf{Gravity multiplet} $\left( g_{\mu\nu}\,, \psi_\mu^+\,, B_{\mu\nu}^+\right) \textbf{:}$ Here $g_{\mu\nu}$ is the metric tensor, $\psi_\mu^+$ is the gravitino, and    $B_{\mu\nu}^+$ is a self-dual 2-form gauge field.
           \item \textbf{Tensor multiplets} $\left(  t\,, B_{\mu\nu}^-\,,  \chi^-  \right)$ \textbf{:} Here  $t$ is a scalar field, $B^-_{\mu\nu}$ is an anti self-dual 2-form gauge field, and $\chi^-$ is a spin $\frac{1}{2}$ Majorana-Weyl fermion of negative chirality. 
         \item \textbf{Vector multiplets} $\left( A_\mu \,,  \lambda^+ \right)$ \textbf{:} Here $A_\mu$ is a gauge field corresponding to the gauge algebra $\mathfrak{g}$, and $\lambda^+$ is a spin $\frac{1}{2}$ adjoint-valued symplectic Majorana-Weyl fermion of positive chirality.
       \item \textbf{Hyper multiplets} $\left( 4 \varphi \,, \psi^-\right) \textbf{:}$ Here $4\varphi$ denotes a pair of complex bosons or four real bosons, and $\psi^-$ is a spin $\frac{1}{2}$ Weyl   fermion of negative chirality.
   \end{itemize}
   We consider six-dimensional $\mathcal{N}=(1,0)$ supergravity theories that contain one gravity multiplet, $n_T$ tensor multiplets, $n_V$ vector multiplets, and $n_H$ hypermultiplets. 
   These type of theories have a gauge group of the form $G = \prod_{\kappa=1}^M G_\kappa \times U(1)^r/\Xi$, where $G_\kappa$ are simple non-abelian gauge group factors,  $\Xi$ is a discrete group and $M\in\mathbb{N}$. 
   Accordingly, the matter content, i.e. hyper- and vector multiplets, transforms under representations of $G$.
   Let us also remark that, whenever $n_T \geq 2$, such a theory has no Lagrangian description~\cite{Taylor:2011wt}.

The quantum anomalies of these theories are encoded in an eight-form anomaly polynomial $I_8(R,F)$, where $R$ denotes the curvature two-form and $F$ the gauge field strength two-forms. These anomalies arise from one-loop box diagrams with four external gauge bosons or gravitons---or a combination of both. The Green--Schwarz mechanism cancels them via tree diagrams mediated by an exchange of $B$ fields, as depicted in Figure~\ref{Fig:1}. 
The six-dimensional gravitational, gauge and mixed gauge-gravitational anomalies cancel when imposing that the anomaly polynomial factorizes as follows~\cite{Green:1984bx,Sagnotti:1992qw,Sadov:1996zm}
\begin{equation}
I_8 \stackrel{!}{=} \frac{1}{2} \Omega_{\alpha \beta}X_4^\alpha\wedge X_4^\beta\,, \quad X_4^\alpha = \frac{1}{2}a^\alpha \mathrm{tr}R\wedge R + 2\sum_\kappa\frac{ b_\kappa^\alpha}{\lambda_\kappa}\mathrm{tr} F_\kappa\wedge F_\kappa + 2b_{ab}^\alpha F^a \wedge F^b \,,
\label{eqn:GS6d}
\end{equation}
where $\Omega_{\alpha\beta}$ is a signature $(1,n_T)$ inner product, 
$a^\alpha$, $b^\alpha_{\kappa}$ and $b_{ab}^\alpha$ are $SO(1,n_T)$ vectors that we call anomaly coefficients, $\lambda_\kappa$ is the Dynkin index of the fundamental representation of the gauge group $G_\kappa$, $F_\kappa$ is the gauge field strength associated to $G_\kappa$ and $F^a$ the gauge field strength associated to an abelian gauge group factor in $G$. 

The generalized Green--Scharz mechanism realized through~(\ref{eqn:GS6d}) imposes a set of conditions that relates the massless spectrum with the anomaly coefficients. Each type of term $R^4,F^2 R^2, \ldots$ in the anomaly polynomial must cancel separately, leading to the anomaly cancellation conditions:~\cite{Park:2011ji} 

\noindent\textbf{Pure gravitational anomaly equations}
\begin{empheq}[box=\fbox]{equation}
\begin{aligned}
R^4      & \;:\; & 273      &= n_H - n_V + 29 n_T\\
(R^2)^2  & \;:\; & a\cdot a &= 9 - n_T
\label{eqn:Grav1}
\end{aligned}
\end{empheq}
\textbf{Mixed anomaly equations}
\begin{empheq}[box=\fbox]{equation}
\begin{aligned}
F_\kappa^2R^2      & \;:\; & a\cdot \frac{b_\kappa}{\lambda_\kappa}      &= \frac{1}{6}\left( A_{\mathbf{Adj}_\kappa} -\sum_{\mathbf{R}}x_{\mathbf{R}} A_{\mathbf{R}} \right)\\
F^a F^bR^2  & \;:\; & a\cdot b_{ab} &= -\frac{1}{6} \sum_{\mathsf{q}_a, \mathsf{q}_b}x_{\mathsf{q}_a,\mathsf{q}_b} \mathsf{q}_{a} \mathsf{q}_{b}
\label{eqn:mixedAn1}
\end{aligned}
\end{empheq}
\textbf{Pure gauge anomaly equations}
\begin{empheq}[box=\fbox]{equation}
\begin{aligned}
F^4      & \;:\; & 0      &=  B_{\mathbf{Adj}_\kappa} -\sum_{\mathbf{R}}x_{\mathbf{R}} B_{\mathbf{R}}\\
(F^2)^2  & \;:\; & \left(\frac{b_\kappa}{\lambda_\kappa} \right)^2 &= \frac{1}{3}\left(\sum_{\mathbf{R} }x_{\mathbf{R}} C_{\mathbf{R}}-C_{\mathbf{Adj}_\kappa}\right) \\
F_\kappa^2 F_\mu^2  & \;:\; & \left(\frac{b_\kappa}{\lambda_\kappa} \right) \cdot \left( \frac{b_\mu}{ \lambda_\mu} \right) &= \sum_{\mathbf{R}, \mathbf{S}} x_{\mathbf{R},\mathbf{S}} A_{\mathbf{R}} A_{\mathbf{S}} \\
(F^i)^3 F_\kappa & \;:\; & 0 &= \sum_{\mathbf{R}, \mathsf{q}_a} x_{\mathbf{R},\mathsf{q}_a} \mathsf{q}_a E_{\mathbf{R}} \\
F^a F^b F_\kappa & \;:\; & \left( \frac{b_\kappa}{\lambda_\kappa} \cdot b_{ab}\right) &= \sum_{\mathbf{R}, \mathsf{q}_a} x_{\mathbf{R},\mathsf{q}_a} \mathsf{q}_a E_{\mathbf{R}} \\
F^a F^b F^c F^d & \;:\; & b_{ab}\cdot b_{cd} + b_{ac}\cdot b_{bd} + b_{ad}\cdot b_{bc} &= \sum_{\mathbf{R}, \mathsf{q}_a} x_{\mathsf{q}_a,\mathsf{q}_b, \mathsf{q}_c \mathsf{q}_d} \mathsf{q}_a \mathsf{q}_b \mathsf{q}_c \mathsf{q}_d \\
\end{aligned}
\label{eqn:pureGauge1}
\end{empheq}
Here $x_{\mathbf{R}}$ denotes the number of hypermultiplets in the representation $\mathbf{R}$ of the gauge group $G_\kappa$,  $x_{\mathbf{R},\mathbf{S}}$ is the number of hypermultiplets in the representation $\mathbf{R}\times \mathbf{S}$ of $G_\kappa\times G_\mu$, $x_{\mathbf{R},\mathsf{q}_a}$ is the number of hypermultiplets in the representation $\mathbf{R}$ of $G_\kappa$ with charge $\mathsf{q}_a$ under $U(1)_a$, $x_{\mathbf{R},\mathsf{q}_a,\mathsf{q}_b}$ is the number of hypermultiplets in the representation $\mathbf{R}$ of $G_\kappa$ with charge $(\mathsf{q}_a,\mathsf{q}_b)$ under $U(1)_a\times U(1)_b$, $x_{\mathsf{q}_a, \mathsf{q}_b}$ is the number of hypermultiplets with charge $(\mathsf{q}_a,\mathsf{q}_b)$ under $U(1)_a \times U(1)_b$ and   $x_{\mathsf{q}_a,\mathsf{q}_b,\mathsf{q}_c,\mathsf{q}_d}$ is the number of hypermultiplets with charge $(\mathsf{q}_a,\mathsf{q}_b,\mathsf{q}_c,\mathsf{q}_d)$ under $U(1)_a\times U(1)_b\times U(1)_c\times U(1)_d$. 
Moreover, we introduced the group theory coefficients $A_{\mathbf{R}}$, $B_\mathbf{R}$, $C_{\mathbf{R}}$ and $E_{\mathbf{R}}$, which are defined through the following relations
\begin{align}
    \mathrm{tr}_{\mathbf{R}} F_\kappa^2 = A_{\mathbf{R}}\mathrm{tr}F_\kappa^2\,, \quad \mathrm{tr}_{\mathbf{R}} F^4 = B_{\mathbf{R}} \mathrm{tr} F^4 + C_{\mathbf{R}}(\mathrm{tr}F^2)^2 \,, \quad \mathrm{tr}_{\mathbf{R}} F^3 = E_{\mathbf{R}} \mathrm{tr} F^3 \,,
    \label{eqn:groupCoef}
    \end{align}
    where $\mathrm{tr}$ denotes the trace in the fundamental representation and $\mathrm{tr}_{\mathbf{R}}$ the trace in the representation $\mathbf{R}$. Let us also note that~(\ref{eqn:pureGauge1}) is meant to be satisfied for $G_\kappa \neq G_\mu$.

\subsection{F-theory picture}
\label{sec:F-theory}

A large class of six-dimensional $\mathcal{N}=(1,0)$ supergravity theories can be realized through F-theory compactifications on an elliptically fibered Calabi--Yau threefold $\pi: X_\star\to B$, such that $B$ is a complex K\"ahler surface. Equivalently, we can describe the fibration in terms of a Weierstrass model
\begin{equation}
y^2 = x^3 + x f+ g \,,
\label{eqn:Weierstrass}
\end{equation}
where $f$ and $g$ are sections of $4\bar{K}_B$ and $6\bar{K}_B$  respectively with $\bar{K}_B$ being the anti-canonical class of $B$.  
F-theory can be interpreted as a compactification of Type IIB string theory on $B$ with a varying axio-dilation identified with the complex structure of the elliptic curve  described by the Weierstrass model~(\ref{eqn:Weierstrass}). See for instance~\cite{Weigand:2018rez,Cvetic:2018bni} for comprehensive reviews on F-theory. On special loci in $B$ the elliptic fiber might degenerate,  giving rise to gauge symmetry enhancements and the appearence of charged matter.

\noindent \textbf{Codimension-one singularities:} A crucial object for an elliptic fibration is the discriminant $\Delta := 4f^3 +27g^2$.  
The divisor $ \{\Delta =0\} \subset B$ 
is physically interpreted as the location of 7-branes that back-react on the axio-dilation, while geometrically it indicates the locus in which the elliptic fiber  pinches. The discriminant locus factorizes into a finite set of irreducible component divisors $\Sigma_\kappa \subset B$ in which the elliptic fibration develops a Kodaira singularity, 
 determined by the vanishing orders of the polynomials $(f,g,\Delta)$. Upon a crepant resolution $\rho : X\to X_\star$, the exceptional fiber over $\Sigma_\kappa$ splits into a chain of rational curves whose topology is that of the affine Dynkin diagram of a Lie algebra $\mathfrak{g}_\kappa$. The resolution introduces a set of exceptional  divisors $E_{I_\kappa}\subset X$, with $I_\kappa =1, \ldots, \mathrm{rk}(\mathfrak{g}_\kappa)$, which are $\mathbb{P}^1$-fibrations over $\Sigma_\mathfrak{\kappa}$ with fibers $C_{{I_\kappa}}$. 
 Together they fulfill the non-trivial intersection relations
 \begin{equation}
 [E_{I_\kappa} ]\cdot [C_{{J_\lambda}}] = - \delta_{\kappa \lambda} C_{IJ,\kappa}\,, \quad [E_{I_\kappa}]\cdot [E_{J_\lambda}]\cdot  \pi^\ast(\alpha)= -\delta_{\kappa \nu }\mathcal{C}_{IJ,\kappa}([\Sigma_{\kappa}] \cdot_B \alpha) 
 \,,
 \end{equation}
 where $\alpha\in H^{1,1}(B)$, $C_{IJ,\kappa}$ and $\mathcal{C}_{IJ,\kappa}$ are the Cartan and coroot intersection matrix of $\mathfrak{g_\kappa}$ respectively. Hence, we relate $E_{I_\kappa}$ and $C_{I_\kappa}$ with the simple coroots and roots of $\mathfrak{g}_\kappa$ respectively.

\noindent \textbf{Codimension-two singularities:} At 
codimension-two loci $\Sigma_\kappa \cap \Sigma_\nu\subset B$, the polynomials $(f,g,\Delta)$ 
might increase their vanishing order.  
This represents an intersection of 7-branes with stretched $(p,q)$-strings that result in localized massless matter. 
 By resolving the codimension-two singularities at points in $B$,  the fiber curves $C_{I_\kappa}$ 
 can split further at $\Sigma_\kappa\cap \Sigma_\nu$ into additional  rational curves 
  $C_{\bm{\lambda}}$, which intersect with the exceptional divisors $E_{I_\kappa}$ as follows
\begin{equation} 
           [C_{\bm{\lambda}}] \cdot [E_{I_\kappa}  ]= \left( \bm{\lambda} , \bm{\alpha}_{I_\kappa}^\vee \right)_{\mathfrak{g}_{\kappa}} = \lambda_{I_\kappa} \in \mathbb{Z}\,.
\label{eqn:DynkinLab}
\end{equation}
We relate the latter intersection numbers with the Dynkin label of a weight $\bm{\lambda}$ in the weight lattice $P(\mathfrak{g}_\kappa \oplus \mathfrak{g}_\nu)$ that defines a weight space $V_{\bm{\lambda}}$ in a highest weight module $V_{\textbf{R}}$,
which has an associated irreducible representation $\textbf{R}$ of the gauge algebra $\mathfrak{g}_\kappa\oplus \mathfrak{g}_\nu$.  
Now, let us consider a curve $C_{\bm{\lambda}_+}$ whose Dynkin label matches with that of the highest weight $\bm{\lambda}_+$ that defines $V_{\mathbf{R}}$. The descendant weights are then realized by taking finite unions of $C_{\bm{\lambda}_+}$ with the fibral curves $C_{I_\kappa}$. In terms of curve classes, 
 we can generate all weights associated to $\textbf{R}$ and establish the geometric-representation correspondence:
\begin{equation}
\label{eqn:Weights}
[C_{\bm{\lambda}}] = [C_{\bm{\lambda}_+}] + \sum_{I_\kappa=1}^{\text{rk}\mathfrak{g}_\kappa} m_{I_\kappa}[ C_{I_\kappa}] + \sum_{J_\nu=1}^{\text{rk}\mathfrak{g}_\nu} n_{J_\nu} [C_{J_\nu} ]
\quad \leftrightarrow \quad
\text{weights $\bm{\lambda}$  in  } V_{\mathbf{R}} = \bigoplus_{\bm{\lambda}} V_{\bm{\lambda}}\,.
\end{equation}
Here the entries $m_{i_I}$ and $n_{j_J}$ are appropriate non-negative integers numbers, 
such that it realizes the Dynkin labels associated with $\bm{\lambda} \in P(\mathfrak{g}_{\kappa} \oplus \mathfrak{g}_{\nu})$.
This way an M2-brane that wraps a curve $C_{\bm{\lambda}}$ realizes a BPS state with weight $\bm{\lambda}$ in the representation $\mathbf{R}$, i.e.  in the weight space $V_{\bm{\lambda}} \subset V_{\mathbf{R}}$.
An anti-M2-brane wrapping the same curve yields a BPS state with negative weight $-\bm{\lambda}$ and hence in the conjugate representation $\bar{\mathbf{R}}$. 
Moreover, the volume of the curve $C_{\bm{\lambda}}$  parametrizes the mass of such BPS states,
which implies that in the zero fiber-volume limit we obtain massless matter states localized at $\Sigma_\kappa \cap \Sigma_\nu\subset B$. 


\noindent\textbf{The Mordell-Weil group:} More generally, an elliptically fibered Calabi-Yau $X$ can contain additional rational sections $s_a$, which are rational maps $s_a: B \rightarrow X$  
that intersect the elliptic fiber in one point, i.e., $\pi \circ s_a = \text{id}_B$. The rational sections $\{s_a\}$ satisfy a group composition law that defines the Mordell-Weil group  $\text{MW}(X)$. 
The Mordell-Weil theorem for function fields states that $\text{MW}(X)$ is a finitely generated abelian group of the form
\begin{equation}
      \text{MW}(X) \cong \mathbb{Z}^{\oplus r} \oplus    \text{MW}(X)_{\text{tor}}. 
\end{equation}
Here the factor $\mathbb{Z}^{\oplus r}$ gives the number of independent rational sections, besides the zero section $s_0$, and $r$ is the rank of the Mordell-Weil group. 
Moreover, $\text{MW}(X)_{\text{tor}}$ denotes the torsional part of the Mordell-Weil group. 
To each rational section in $\text{MW}(X)$ there is a divisor $S_a \equiv \text{div}(s_a)$. 
However, the rational section divisors $\{S_{a}\}$ do not follow the additive group composition law of $\text{MW}(X)$.  
Instead, 
a set of divisors satisfying such a property is given by the Shioda map $\sigma: \mathrm{MW}(X) \to \mathrm{NS}(X) \otimes \mathbb{Q}$, defined as follows
\begin{equation}
\label{eqn:Shioda}
\sigma(s_a) = S_a - S_0 -\pi^{-1}\left( \pi_* \left( \left(S_a-S_0\right)\cdot S_0 \right)\right) 
+ \sum_{\kappa} \left(S_a\cdot C_{I_\kappa} \right) \mathcal{C}^{IJ,\kappa} E_{J_\kappa}\,.
\end{equation}
Here $\mathcal{C}^{IJ,\kappa}$ is the inverse of the Cartan matrix associated to the Lie algebra $\mathfrak{g}_\kappa$.

Let us remark that in F-theory the information about the global structure for a non-abelian gauge group $G$ with Lie algebra $\mathfrak{g}$ is reflected in the representation theory content of the spectrum. For a torsional section $s_t$, its image under the Shioda map is trivial. This gives rise to integrality constraints for the Dynkin labels for a given matter curve  $C_{\bm{\lambda}}$, thereby restricting the possible matter representations. 
Globally, this leads to a non-simply-connected gauge group in which $\pi_1(G)_{\mathrm{tor}} \cong \mathrm{MW}(X)_{\mathrm{tor}}$.\footnote{The case for gauge groups of the form $G\times U(1)^r$ is more subtle but expected to be similar. See~\cite{Cvetic:2017epq} for further discussion on this point.}

\noindent \textbf{Anomaly coefficients:} Let us now relate the intersection theory of $\pi : X \to B$ to the anomaly coefficients of the associated effective field theory in six dimensions reviewed in section~\ref{sec:Anomalies}.  
The number of tensor multiplets is given by $n_T = h^{1,1}(B) -1$, as shown in~\cite{Vafa:1996xn}. 
Geometrically, the $SO(1,n_T)$ vectors $a^\alpha$, $b_{IJ,\kappa}^\alpha$ and $b_{ab}^\alpha$ are determined by~\cite{Park:2011ji} 
\begin{align}
    a^\alpha = -[\bar{K}_B] \cdot C^\alpha\,,\quad b_\kappa^\alpha = [\Sigma_\kappa]\cdot C^\alpha\,,  \quad b_{ab}^\alpha = -\pi_\ast (\sigma(s_a) \cdot \sigma(s_b))\cdot C^\alpha\,,
    \label{eqn:AnCoef}
\end{align}
where $\{C^\alpha\}$ is a given basis of curve classes for $H_2(B,\mathbb{Z})$. The $SO(1,n_T)$ metric can then be read off by inverting
\begin{equation}
 \Omega_{\alpha \beta} = (\Omega^{\alpha \beta})^{-1} \,, \quad \Omega^{\alpha \beta} = C^\alpha \cdot C^\beta   \,.
\end{equation}


\noindent \textbf{Massless hypermultiplets and Kaluza-Klein towers:} In M-theory compactified on a smooth elliptic fibration $\pi: X \to B$, massless abelian vector fields emerge in the five-dimensional effective action from the expansion of the three-form $C_3$ along a basis of harmonic two-forms. By the Shioda--Tate--Wazir theorem, it can be expanded as 
\begin{equation}
 C_3 =\hat{A}^0\wedge  [\hat{S}_0]+ A^{I_\kappa} [E_{I_\kappa} ]+A^{a}\sigma(s_a)+ A^\alpha \wedge [D_\alpha]\,,
\end{equation}
where $\{D_\alpha\}$ is a basis of pullback divisors $D_\alpha = \pi^\ast D_\alpha'$, with $\{D_\alpha'\}$ a basis of $H^{1,1}(B)$. Moreover, we also introduced the shifted zero section divisor
\begin{equation}
   [ \hat{S}_0 ]: = [S_0] +\frac{1}{2} \pi^\ast [\bar{K}_B]\,.
    \label{eqn:KKDiv}
\end{equation}
M2-branes wrapping holomorphic curves $C \subset X$ give rise to  BPS particles in five dimensions with mass $m = \int_C \omega$, where $\omega$ denotes the K\"ahler form of $X$.
We will be particularly interested in curves whose classes lie in the relative Mori cone $\mathrm{Mori}(X/B)$, which is defined as the set of curve classes $[C]\in \mathrm{Mori}(X)$ such that $\pi_\ast [C] =0$. 
Such particles carry the following types of abelian charges
\begin{equation}
n = [\hat{S}_0] \cdot [C]\,, \quad  \lambda_{I_\kappa}= [E_{I_\kappa}]\cdot [C] \,, \quad\mathsf{q}_a = [\sigma(s_a) ]\cdot[ C ]\,. 
\end{equation}
Here $n$ is the Kaluza-Klein charge, since $\hat{A}^0$ is interpreted as the Kaluza--Klein gauge field due to $S^1$ compactification of F-theory~\cite{Park:2011ji}. The charges 
$\lambda_{I_\kappa}$ are weights under the Cartan subalgebra $\mathfrak{h}_\kappa \subset \mathfrak{g}_\kappa$ associated with the vector fields $A^{I_\kappa}$, while $\mathsf{q}_a$ denote the abelian charges stemming from a $U(1)_a$ factor of F-theory. 


The six-dimensional massless spectrum of F-theory is thus identified with the zero-th mode sector of these Kaluza--Klein states arising from curves in $\mathrm{Mori}(X/B)$, upon taking the zero volume limit for the elliptic fiber. In the next chapter, we will relate the enumerative geometry of this type of curves---together with their Kaluza--Klein modes---to the multiplicities of massless fields in six dimensions.

\noindent\textbf{Discrete gauge symmetries:} F-theory can also be defined on Calabi--Yau manifolds $X$ that are torus fibered over a K\"ahler base $B$ but admit no section, i.e. they are not elliptically fibered. Instead, they possess multi-sections or $N$-sections. Compactifications of this type are known as genus one fibered Calabi--Yau manifolds and they provide a geometric realization of discrete gauge symmetries $\mathbb{Z}_N$ in their corresponding effective field theories~\cite{Braun:2014oya,Morrison:2014era}. 
Alternatively, these models can be studied via their associated Jacobian fibration $J(X)$, which admits a section and a Weierstra\ss{}   model description at the cost of introducing $\mathbb{Q}$-factorial terminal singularities~\cite{Braun:2014oya}. We will not discuss this class of fibrations further and instead refer the reader to the literature for more details~\cite{Cvetic:2018bni,Duque:2025kaa}.

\section{Modularity of Calabi-Yau threefolds}
\label{sec:3}

In this chapter we introduce our main computational object that carries both the classical intersection data and the Gromov--Witten invariants of Calabi--Yau manifolds. Our primary motivation is to count curves on elliptically fibered Calabi--Yau threefolds, restricted to fiber classes, in order to determine the massless spectrum of six-dimensional theories. This counting problem exhibits a rich modular structure because the elliptic fiber induces 
an $\mathrm{SL}(2,\mathbb{Z})$ action that will play a central role in subsequent chapters. To prepare for this, we review basic aspects of topological string theory and summarize tools from homological mirror symmetry that allow us to describe the modular group action on the relevant curve classes explicitly.

\subsection{Topological string theory three-point couplings}


In this section we briefly introduce our main object of interest, the  three-point correlation functions, which are central observables in both the A- and B-model topological string theories. A convenient method for computing these observables involves using period integrals on the B-model side, followed by the application of mirror symmetry to obtain the A-model counterpart. Let us point out that on the A-model side the three-point correlation functions admit an enumerative geometric interpretation, which will play a central role in the subsequent sections of this work.

The B-model is associated with the study of complex structure deformations of a Calabi--Yau threefold $Y$. These deformations are parameterized by the complex structure moduli space $\mathcal{M}_{\mathrm{cs}}(Y)$, which is unobstructed by the Tian--Todorov theorem and has complex dimension $h:=h^{2,1}(Y)$. The geometry of $\mathcal{M}_{\mathrm{cs}}(Y)$ can be encoded in the variation of the nowhere-vanishing holomorphic $(3,0)$-form $\Omega(\bm{u})$, where $\bm{u} \in \mathcal{M}_{\mathrm{cs}}(Y)$, and in its period integrals over a suitable choice of cycles in $H_3(Y,\mathbb{Z})$.  
A convenient choice is given by the symplectic basis 
of cycles $A^I, B_J \in H_3(Y,\mathbb{Z})$ and a dual basis in cohomology $\alpha_I, \beta^I \in H^3(X,\mathbb{Z})$ with $I=0, \ldots , h$, 
such that 
\begin{equation}
A^J\cap B_I = -B_I\cap A^J = \int_{Y} \alpha_I\wedge\beta^J =- \int_Y \beta^J \wedge \alpha_I = \int_{A^J} \alpha_I = \int_{B_I} \beta^J = \delta_I^J\,.
\end{equation}
Accordingly, we define the period integrals in the symplectic basis as follows 
\begin{equation}
\vec{\Pi} = (F_I, X^I)^t\,, \quad X^I(\bm{u}) = \int_{A^I} \Omega(\bm{u}) \,, \quad F_I(\bm{u}) = \int_{B_I}\Omega(\bm{u}) \,. 
\label{eqn:sympBasis}
\end{equation}
By the local Torelli theorem, the periods $X^I$ constitute local homogeneous coordinates  
for $\mathcal{M}_{\mathrm{cs}}(Y)$ due to   
possible redefinitions of $\Omega$ by a non-vanishing scaling factor. 
However, a choice can be made for a period $X^0$ 
to define local inhomogeneous coordinates for $\mathcal{M}_{\mathrm{cs}}(Y)$ given by the ratios
\begin{equation}
t^i(\bm{u}) = \frac{X^a(\bm{u})}{X^0(\bm{u})}\,, \quad i =1,\ldots , h\,,
\label{eqn:tRat}
\end{equation}
such that $X^0$ stays regular on a given patch of the moduli space. Notice that the ratio in~(\ref{eqn:tRat}) is then locally invertible for $\bm{z}$ in term of the coordinates $\bm{t}$.  

The nowhere vanishing holomorphic form $\Omega$ is a section of a holomorphic line bundle $\mathcal{L}$ over $\mathcal{M}_{\mathrm{cs}}(Y)$ that transforms as $\Omega(\bm{z}) \to e^{f(\bm{z})} \Omega(\bm{z})$ with $f$ being a holomorphic function. 
The three-point couplings on the B-model side have the geometrical definition
\begin{equation}
C_{ijk}(\bm{u})= \int_Y \Omega \wedge \partial_{u^i} \partial_{u^j} \partial_{u^k} \Omega \,,
\end{equation}
where $u^i$ are any parametric choice of complex structure coordinates.  
Moreover, $C_{ijk}$ is a section of $\mathrm{Sym}_3(T^\ast_{1,0}(\mathcal{M}_{\mathrm{cs}})) \otimes \mathcal{L}^2$. 

Mirror symmetry exchanges the complex structure and the complexified K\"ahler structure, and in turn, translates the information from the B-model to the A-model via the mirror map $t: \mathcal{M}_{\mathrm{cs}}(Y) \to \mathcal{M}_{\mathrm{ks}}(X)$ which is defined patchwise as an inversion of convergent power series.  Thus, the three-point  couplings on the A-model side can be obtained from the B-model three-point coupling by performing the transformation
\begin{equation}
C_{ijk}(\bm{t}) = \frac{1}{(X^0)^2} \frac{\partial u^l}{\partial t^i}\frac{\partial u^m}{\partial t^j}\frac{\partial u^n}{\partial t^k} C_{lmn}(\bm{z} \circ \bm{t})\,,
\label{eqn:3ptTrans}
\end{equation}
where  $t^i$ are the flat coordinates~(\ref{eqn:tRat}). 

In this work, we are interested in studying the A-model three-point couplings in the vicinity of the large volume region $\mathrm{Im}(t^i) \to \infty$ in the stringy K\"ahler moduli space. 
 There, the three-point copuplings  
 receive the interpretation of a quantum cohomology deformation of the classical intersections as follows
\begin{equation}
C_{ijk}(t) = \int_{X}\omega_i \wedge \omega_j \wedge \omega_k  + \sum_{\beta \in H_2(X,\mathbb{Z})}\langle\omega_i ,\omega_j, \omega_k \rangle_\beta Q^\beta\,,
\end{equation}
where $\{\omega_i\}$ is a basis for $H^\ast(X)$, $Q^\beta =\exp(2\pi i \bm{t }\cdot \beta)$ and $\langle \omega_i,\omega_i,\omega_k\rangle_\beta$ are determined by Gromov--Witten invariants~\cite{Cox:2000vi}. 
See Appendix~\ref{App:GW} for its definitions. 
Alternatively, we can describe the three-point coupling in terms of triple derivatives $C_{ijk}(\bm{t}) = \partial_{t^i} \partial_{t^j} \partial_{t^k} F(\bm{t})$, 
where $F(\bm{t})$ is the inhomogeneous holomorphic prepotential, which receives instanton corrections in terms of Gopakumar-Vafa invariants at large volume~\cite{Gopakumar:1998ii,Gopakumar:1998jq}.



\noindent\textbf{Monodromy transformations:} The B-model periods transform non-trivially under monodromies when taking parallel transport along loops that encircle singular loci in the complex structure moduli space. Such transformations act on the period vectors $\vec{\Pi}$ as
\begin{equation}
\vec{\Pi} \mapsto M \vec{\Pi}\,, \quad M \in \Gamma_Y \subset \mathrm{Sp}(2h^{2,1}(Y) +2,\mathbb{Z})\,,
\label{eqn:monBmod}
\end{equation}
where $\Gamma_Y$ denotes the monodromy group.  Consequently,~(\ref{eqn:3ptTrans}) implies that the three-point couplings transform covariantly under the induced action of a given monodromy transformation. 

For our purposes, it is relevant to consider a universal set of monodromies of elliptically fibered Calabi--Yau threefolds, whose induce action on the moduli can be identified with an $\mathrm{SL}(2,\mathbb{Z})$ modular group action associated with the elliptic fiber. To that end, we next invoke the machinery of homological mirror symmetry.

\subsection{Homological mirror symmetry and brane charges}
\label{sec:HMS}


The homological mirror symmetry conjecture proposes the equivalence of categories $D^b(X) \cong D^b\mathcal{F}(Y)$, 
if and only if $(X,Y)$ are Calabi-Yau mirror dual pairs~\cite{MR1403918}. Here $D^b(X)$ is the bounded derived category of coherent sheaves, whereas $D^b\mathcal{F}(Y)$ is a derived version of the Fukaya 
category. Before we
explain the objects of each category---topological branes---let us establish their connection with our motivations concerning monodromy transformations on periods. 

Homological mirror symmetry provides a powerful method that furnishes an integral basis of periods by considering central charges of branes. 
The upshot is that periods $\Pi_I$ represent a basis for the lattice of central charges of D-branes $\Lambda_{\text{D}}$. 
The central charge of a D-brane follows from the formula $Z(\gamma) =  \gamma^I \Pi_I $, where $\gamma^I \in \mathbb{Z}$;
 a D-brane is subjected to the BPS mass bound $M \geq \vert Z(\gamma) \vert$.  
 When branes saturate the BPS bound, they are said to be topological. 
 There are two types of topological D-branes, 
 which exchange under mirror duality, 
 corresponding to objects in the categories $D^b(X)$ and $D^b\mathcal{F}(Y)$.
 We describe such objects briefly next.
\\\\
\textbf{A-branes:} 
The objects associated with $D^b\mathcal{F}(Y)$ are topological A-branes, which are characterized by calibrated Lagrangian three-cycles $L \subset Y$ endowed with a flat $U(1)$ connection. 
The central charge of topological A-branes reads from the period integral~\cite{MR2567952}
\begin{equation}
\Pi_{\mathrm{A}}(L) = \int_L \Omega \,,
\end{equation}
where $\Omega$ denotes the holomorphic $(3,0)$-form in $Y$. For a more precise mathematical description of $D^b\mathcal{F}(Y)$ and further discussions of A-branes, we refer the reader to~\cite{MR2567952}. 

\noindent\textbf{B-branes:} As proposed in~\cite{MR1403918}, 
the topological B-branes $\mathcal{E}^\bullet$ on $X$ are objects in the derived category of bounded complexes of coherent sheaves $D^b(X)$.  
These are described as bounded complexes of locally free sheaves
\begin{equation}
\mathcal{E}^\bullet :0 \rightarrow \cdots  \rightarrow \mathcal{E}^{i-1} \rightarrow \mathcal{E}^i \rightarrow \mathcal{E}^{i+1}\rightarrow \cdots \rightarrow 0 \,.
\end{equation}
Each locally free sheaf $\mathcal{E}^i$ is equivalent to a vector bundle $E^i$. 
Physically, we interpret $\mathcal{E}^{2i}$ as coincident branes, while $\mathcal{E}^{2i+1}$ as coincident anti-branes~\cite{MR1726156,MR2567952}. 
The object that measures  the RR charges of this configuration of branes is the $K$-theory group $K(X)$ \cite{Witten:1998cd}. The latter is defined by the set of complex vector bundles $(E,F)$, which are subjected to an equivalence relation $(E,F) \sim (E\oplus H, F \oplus H)$ for any vector bundle $H$.
Let us introduce the Chern character ring homomorphism $\text{ch} : K(X) \rightarrow H^{2*}(X,\mathbb{Q})$, which acts on the associated vector bundles
\begin{equation}
\text{ch}(\mathcal{E}^\bullet) = \cdots - \text{ch}\left(E^{2i-1}\right) + \text{ch}\left(E^{2i}\right)-\text{ch}\left(E^{2i+1}\right) +\cdots
\end{equation}
In the large volume limit $\mathrm{Im}(t^i) \to \infty$, we assign to each B-brane a central charge whose leading behaviour is given by the asymptotic period formula
\begin{equation}
\label{eqn:Bcentral}
\Pi_{\mathrm{B}}\left( \mathcal{E}^\bullet \right) = \int_X e^{\omega} \Gamma_{\mathbb{C}}(X) \text{ch}  (\mathcal{E}^\bullet)^\vee + \mathcal{O}(e^{2\pi i \bm{t}})\,,
\end{equation}
where $\omega$ denotes the complexified K\"ahler form of $X$, the exponentially suppressed terms  correspond to worldsheet instanton corrections,  the Gamma class of $X$ is
\begin{equation}
 \Gamma_{\mathbb{C}}(X)= 1 + \frac{1}{24}c_2(X) + \frac{\zeta(3)}{(2\pi i)^3} c_3(X)
\end{equation}
and the operator $\vee: H^{2*}(X) \rightarrow H^{2*}(X)$ acts as $\gamma^\vee \mapsto (-1)^k \gamma$ for $\gamma \in H^{k,k}(X)$. 

We now introduce a universal basis for B-branes whose associated periods span the lattice of central charges at the large volume limit in the stringy K\"ahler moduli space of $X$~\cite{Gerhardus:2016iot}. First, 
we have the six-brane structure sheaf $\mathcal{O}_X$ and the zero-brane given by the skycraper sheaf $\mathcal{O}_p$ with support on a point $p \in X$. 
Moreover, a collection of four-branes with support on effective divisors $D\subset X$ are given by the complexes 
\begin{equation}
\mathcal{E}_D^\bullet: 0 \longrightarrow \mathcal{O}_X(-D)  \longrightarrow \mathcal{O}_X \longrightarrow 0\,.
\end{equation}
Lastly, a set of 2-branes with support on curves $C \subset X$ are given by
\begin{equation}
\mathcal{C}^\bullet = \iota_{!}\mathcal{O}_{C}(K_{C}^{1/2})\,,
\end{equation}
where we introduced the embedding $\iota : C \hookrightarrow X$, its K-theoretic push-forward $\iota_!: K(C) \to K(X)$ and $\mathcal{O}_{C}(K_{C}^{1/2})$ is the structure sheaf twisted by a spin bundle $K_{C}^{1/2}$ of $C$. This construction of even branes by~\cite{Gerhardus:2016iot} furnishes a complete integral basis of periods as we outline next. 

Let us consider a pair of dual bases for the complexified K\"ahler form $\omega = t^i \omega_i$. Here $\{\omega_i\}$ is a basis of divisor clases for $H^{1,1}(X)$ and $t^i = \int_{C^i}\omega$, where the curve classes $\{[C^i]\}$ form a basis for the Mori cone $\mathrm{Mori}(X)$. Using~(\ref{eqn:Bcentral}) together with the aforementioned basis of B-branes, one finds 
the asymptotic expressions
\begin{align}
\label{eqn:Qperiods}
\begin{split}
\Pi_X &= \frac{1}{6} c_{ijk} t^i t^j t^k + c_i t^i + \frac{\zeta(3)}{(2\pi i )^3}\chi(X) + \mathcal{O}(e^{2\pi i \bm{t}}) \,, \\
\Pi_{\omega_i} & = - \frac{1}{2} c_{ijk} t^j t^k - \frac{1}{2}c_{iij}t^j -\frac{1}{6}c_{iii} -c_i +\mathcal{O}(e^{2\pi i \bm{t}})\,, \\
\Pi_{C^i} & = t^i +\mathcal{O}(e^{2\pi i \bm{t}}) \,, \\
\Pi_{\mathrm{pt}} &= -1+\mathcal{O}(e^{2\pi i \bm{t}}) \,.
\end{split}
\end{align}
Here we introduced the shorter notation $\Pi_V$ to denote the B-brane periods $\Pi_\mathrm{B}(\mathcal{E}^\bullet)$ associated to a B-brane $\mathcal{E}^\bullet$  supported in a subvariety $V\subseteq X$, in accordance to the basis of B-branes we introduced from~\cite{Gerhardus:2016iot}. Moreover, 
$\chi(X)$ is the Euler characteristic of the Calabi-Yau threefold, 
\begin{equation}
c_{ijk} = \int_{X} \omega_i\wedge \omega_j \wedge \omega_k\,, \qquad c_i = \frac{1}{24} \int_{X} c_2(X)\wedge \omega_i\,
\end{equation}
and we also introduced the following intersection relations 
\begin{equation}
c_{iij} =\int_X \omega_i \wedge \omega_i \wedge \omega_j \,, \quad c_{iii} = \int_X \omega_i \wedge \omega_i \wedge \omega_i \,,
\end{equation}
for which the Einstein summation convention does not apply. Note that basis choice in~(\ref{eqn:Qperiods}) defines a vector period
\begin{equation}
\vec{\Pi}(\bm{t}) = \Big( \Pi_X , \Pi_{\omega_i} ,\Pi_{C^i}, \Pi_{\mathrm{pt}}\Big) \,,
\label{eqn:basisPer}
\end{equation}
which can be brought into the symplectic basis form~(\ref{eqn:sympBasis}) by an appropriate linear transformation.

\noindent\textbf{Stability conditions and BPS invariants: } The objects of $D^b(X)$ are subject to stability conditions determined by their central charge $Z : \Lambda_D \to \mathbb{C}$, which varies holomorphically over the stringy K\"ahler moduli space of $X$. 
We say that an object $\mathcal{E}^\bullet$ with charge $\gamma \in \Lambda_D$ is semistable if for every subobject $\mathcal{F}^\bullet \subset \mathcal{E}^\bullet$ with charge $\gamma'\in\Lambda_D$ we have  $\varphi (\gamma') \leq \varphi(\gamma)$, where $\varphi(\gamma)$ is the argument of the central charge $Z(\gamma)$; we call $\mathcal{E}^\bullet$ stable if $\varphi(\gamma') < \varphi(\gamma)$ for all such subobjects.  Physically, the Donaldson--Thomas invariants $\Omega(\gamma)$ correspond to protected BPS indices  given by angular momentum weighted traces over BPS states with charge $\gamma$~\cite{Denef:2007vg,Kontsevich:2008fj}.\footnote{In this picture we are considering BPS particles realized as bound states of even D-branes wrapping even cycles in Type IIA on $X$.} They are locally constant functions of the K\"ahler moduli $t^i$ through their dependence on $Z(\gamma)$ and are robust under complex structure deformations of $X$. 

Mathematically, the physical notion of stability for topological branes is formalized through Bridgeland stability conditions,  defined by a set of axioms in~\cite{Bridgeland:2006bzr}, which  specialize to $\Pi$-stability conditions in our physics discussion.   
Accordingly, it is possible to define generalized Donaldson--Thomas invariants as the Euler characteristic of moduli spaces of semistable objects in a triangulated category~\cite{Joyce:2008pc}, such as $D^b(X)$. 
See~\cite{Alexandrov:2023zjb} for a thorough, physics oriented guide to Bridgeland stability conditions and generalized Donaldson--Thomas invariants, together with impressive applications in topological string theory~\cite{Alexandrov:2023zjb,Alexandrov:2023ltz}.

Let us remark that $\Omega(\gamma)$ may jump
whenever a stable object $\mathcal{E}^\bullet \in D^b(X)$ with charge $\gamma$ becomes unstable. This can happen when the central charge $Z(\gamma')$ of a sub-object $\mathcal{F}^\bullet \subset \mathcal{E}^\bullet$ with charge $\gamma'$ becomes aligned with the central charge $Z(\gamma)$ of $\mathcal{E}^\bullet$, i.e. when $\varphi(\gamma') = \varphi(\gamma)$.  
This condition defines a wall of marginal stability, which is a codimension-one locus in the stringy K\"ahler moduli space of $X$. The discontinuities of the Donaldson--Thomas invariants upon crossing such walls in the moduli space are described through the Kontsevich--Soibelman wall-crossing formula~\cite{Kontsevich:2008fj}. Physically, these jumps are due to the decay of BPS bound states into smaller stable constituents. 

What is important for our purposes is that, away from wall-crossing phenomena, the Donaldson-Thomas invariants are invariant under monodromy transformations. In other words
\begin{equation}
\Omega(M \cdot \gamma; M\cdot \bm{t}) = \Omega(\gamma;\bm{t}) \,.
\label{eqn:BPSInd}
\end{equation}
Here $M$ is a symplectic transformation in $\mathrm{Sp}(2h^{1,1}(X)+2,\mathbb{Z})$ that acts on $\Lambda_D$ and on the complexified K\"ahler moduli $t^i$ through period transformations, as described in the mirror dual picture on the mirror Calabi--Yau $Y$ in~(\ref{eqn:monBmod}).

\subsection{The Fourier--Mukai transform and monodromies}

The homological mirror symmetry conjecture also proposes that monodromy transformations of a given Calabi--Yau manifold $X$ lift to auto-equivalences in $D^b(X)$. 
In particular, when $X$ admits an elliptic fibration, there is a universal set of auto-equivalences of $D^b(X)$ whose induced action on the K\"ahler moduli can be identified with transformations of
the full modular group $\mathrm{SL}(2, \mathbb{Z})$~\cite{Andreas:2000sj,Andreas:2001ve,Schimannek:2019ijf}. Similarly,  when $X$ admits a genus one fibration, a subset of auto-equivalences of $D^b(X)$ gives rise to an action of a congruence subgroup
$\Gamma \subset \mathrm{SL}(2, \mathbb{Z})$~\cite{Cota:2019cjx,Knapp:2021vkm}.  
Let us now discuss these types of transformations in more detail.  

 Let us consider a pair of smooth projective varieties $X_1$ and $X_2$. A theorem by Orlov states that the equivalences of $D^b(X_1)$ and $D^b(X_2)$ are always expressible as a Fourier--Mukai transformation
\begin{equation} 
\Phi_{\mathcal{E}}:\mathcal{F}^\bullet \mapsto R\pi_{1, \ast}\left(\mathcal{E}\otimes_L L\pi_2^\ast\mathcal{F}^\bullet\right)\,,
\end{equation}
where $\pi_i$ is the projection from $X_1 \times X_2$ to $X_i$ and $\mathcal{E}$ is the so-called Fourier--Mukai kernel, which is a quasicoherent sheaf in $D^b(X_1\times X_2)$. 
Here $L$ and $R$ indicate the derived version of a functor to be taken~\cite{Andreas:2004uf}. Note that auto-equivalences correspond to the case in which $X_1 = X_2$. Ultimately, we are interested in how a given Fourier--Mukai transform $\Phi_{\mathcal{E}}: D^b(X) \to D^b(X)$ acts on B-brane central charges
, i.e. we aim to determine the monodromy induced by that Fourier--Mukai transformation.

As a first example, we regard the large volume monodromy transformations in $\mathcal{M}_{\mathrm{ks}}(X)$ and their lifts to auto-equivalences of $D^b(X)$. Let $D \subset X$ be a divisor and consider the Fourier--Mukai kernel $\mathcal{E} = \mathcal{O}_\Delta(D)$ with $\Delta$ being the diagonal in $X\times X$. The corresponding Fourier--Mukai transform acts on an object as twisting by a line bundle~\cite{Andreas:2004uf}
\begin{equation}
    \Phi_\mathcal{E}: \mathcal{F}^\bullet \mapsto \mathcal{F}^\bullet \otimes \mathcal{O}(D)\,.
\end{equation}    
    The induced monodromy transformations can then be read off from the B-brane periods~(\ref{eqn:Bcentral}) of the transformed objects $\Phi_{\mathcal{E}}(\mathcal{F}^\bullet)$. 
In particular, for a colletion of divisors $D_i \subset X$ with $[D_i] = \omega_i$, this procedure yields monodromies $M_i$ acting on the K\"ahler moduli by $M_i : t^j \mapsto t^j + \delta_i^j \,$~\cite[(3.17)]{Cota:2019cjx}.

In the case that we have a smooth torus fibration $\pi : X \to B$, a crucial Fourier--Mukai kernel is the ideal sheaf of the relative diagonal $\mathcal{I}_{\Delta_B}$ in the relative fiber product $X\times_B X$, whose corresponding transform we call the \textit{relative conifold transformation}~\cite{Schimannek:2019ijf,Cota:2019cjx}. It yields the following action on the corresponding B-brane charges of $\mathcal{F}^\bullet$
\begin{equation}
\label{eqn:RCT}
\mathrm{ch}\left(\Phi_{\mathcal{I}_{\Delta_B}}(\mathcal{F}^\bullet) \right) = \mathrm{ch}(\mathcal{F}^\bullet) - \pi_{2,\ast}\left[\pi_1^\ast \left(\mathrm{ch}(\mathcal{F}^\bullet) \mathrm{Td}(T_{X/B})\right)\right]\,,
\end{equation}
where $\mathrm{Td}(T_{X/B})$ is the Todd class of the virtual relative tangent bundle. 
The respective monodromy for the relative conifold transformation can be obtained upon insertion of the transformed B-brane Chern character~(\ref{eqn:RCT}) into the B-brane period formula~(\ref{eqn:Bcentral}). 
See for instance~\cite[Section 2.1]{Schimannek:2019ijf} for a guide on how to compute~(\ref{eqn:RCT}).

For the following discussion, we consider smooth torus fibered Calabi--Yau threefolds $\pi : X \to B$  containing a zero $N$-section divisor $S_0$ with $N\geq 1$, a collection of rational $N$-section divisors $S_a$, resolution divisors $E_{I_\kappa}$ due to ADE singularities fibered over curves $\Sigma_\kappa \subset B$,   and  pullback divisors $D_\alpha = \pi^\ast D'_\alpha$, where $\{D_\alpha'\}$ forms a basis of divisors in $H^{1,1}(B)$. 
Note that this set of divisors is simply a generalization of those introduced in section~\ref{sec:F-theory}, which correspond to the case $N=1$. Then, 
by the Shioda--Tate--Wazir theorem 
we can expand the K\"ahler form as follows~\cite{Cota:2019cjx}
\begin{equation}
\omega = \hat{S}_0 \tau + \sigma(S_a) z^a+E_{I_\kappa} z^{I_\kappa} +   D_\alpha \hat{t}^\alpha \,. 
\label{eqn:KmodBasis}
\end{equation}
Here we introduced the object~\cite{Cota:2019cjx,Pioline:2025uov} 
\begin{equation}
\hat{S}_0 = S_0 -\frac{1}{2N} \pi^\ast \pi_\ast(S_0 \cdot S_0) \,,
\end{equation}
which specializes to~(\ref{eqn:KKDiv}) for $N=1$. Moreover, $\sigma(S_a)$ denotes the Shioda map associated to a rational $N$-section $S_a$, which extends to genus one fibrations and is given by~(\ref{eqn:Shioda}).  
Lastly, $\tau$ is the volume of a curve $C_\tau$ whose class is related to the generic torus fiber class $[\mathbb{E}_\tau]$ by $[\mathbb{E}_\tau] = N [C_\tau]$, $z^a$ and $z^{I_\kappa}$ are the volumes of some isolated fibral curves in $X$, and 
\begin{equation}
    \hat{t}^\alpha = t^\alpha +\frac{1}{2N} \ell^\alpha \tau\,, \quad \ell^\alpha = -\Omega^{\alpha \beta} S_0 \cdot S_0 \cdot D_\alpha\,,
\end{equation}
where $t^\alpha$ is the volume of a base curve whose class is given by $ C^\alpha =\Omega^{\alpha \beta} (S_0 \cdot D_\beta)/N$. In what follows, it will be convenient to introduce the shorthand $z^A \in\{z^a,z^{I_\kappa}\}$.

Let $U$ be the monodromy associated to the relative conifold transformation. 
We are interested in the sub-action of such a monodromy due to two-branes with support on curves in $X$ and the zero-brane with support on a point in $X$. The reason for this is that their corresponding period ratios allow us to define local coordinates in the stringy K\"ahler moduli space of $X$, i.e. 
\begin{equation}
    \tau =  -\Pi_{C_\tau}/\Pi_{\mathrm{pt}}\,, \quad z^A = -\Pi_{C^A}/{\Pi_{\mathrm{pt}}} \,, \quad t^\alpha = -\Pi_{C^\alpha}/{\Pi_{\mathrm{pt}}}\,,
\end{equation} 
where $C^A$ denotes an isolated fibral curve in $X$ whose complex K\"ahler volume is $z^A$. 
With this in mind, we calculate~(\ref{eqn:RCT}) and find the following action on the periods
associated to zero- and two-branes~\cite{Schimannek:2019ijf,Cota:2019cjx}
\begin{align}
\begin{split}
U:\begin{cases}
\Pi_{\mathrm{pt}}\mapsto \Pi_{\mathrm{pt}} - N\Pi_{C_\tau} \,, \quad\\
\Pi_{C_f}\mapsto \Pi_{C_f} \,, \\ 
\Pi_{C^\alpha}  \mapsto  \Pi_{C^\alpha} - \Omega^{\alpha \beta} \Pi_{D_\beta} - \frac{1}{2} \Omega^{\alpha \beta}(\Omega_{\beta \beta}-a_\beta)\Pi_{C_\tau}\,.
\end{cases}
\end{split}
\label{eqn:Usub}
\end{align}
Here $C_f$ denotes a curve in $X$ whose class is in $\mathrm{Mori}(X/B)$ and $a_\alpha = c_1(B) \cdot D_\alpha'$. 
In terms of K\"ahler coordinates, the relative conifold monodromy acts as follows~\cite{Schimannek:2019ijf,Cota:2019cjx}
\begin{align}
\begin{split}
U:
\begin{cases}& \tau \mapsto \frac{\tau}{1+N\tau}\,, \\ 
&z^A \mapsto \frac{z^A}{1+N\tau} \,,
\\
& \hat{t}^\alpha \mapsto \hat{t}^\alpha+ \frac{c^\alpha}{2} -\frac{N}{1+N\tau}
b_{AB}^{\alpha}z^A z^B + \mathcal{O}(e^{2\pi i \hat{t}^\alpha})
\,.
\end{cases}
\label{eqn:Umoduli}
\end{split}
\end{align}
Here $c^\alpha =\Omega^{\alpha\beta} (c_2(T_X) \cdot D_\beta)/12$ and $b_{AB}^\alpha = \Omega^{\alpha \beta} (b_{AB} \cdot D_\beta') $, where we also introduced the shorthand convention
\begin{equation}
b_{AB} = \begin{cases} 
 -\frac{1}{N} \pi_\ast (\sigma(S_a) \cdot \cdot\sigma(S_b)) &  \text{for } A=a, B =b\,,\\
 - \frac{1}{N}\pi_\ast(E_{I_\kappa}\cdot E_{J_\kappa}) & \text{for } A={I_\kappa}, B= J_\kappa\,, \\
 0 & \text{otherwise}\,.
\end{cases}
\end{equation}
Note that the transformation~(\ref{eqn:Umoduli}) of $\tau$ and $z^A$ is exact, whereas the transformation of $\hat{t}^\alpha$ receives exponentially suppressed corrections. With this choice of coordinates, we can relate the action~(\ref{eqn:Umoduli}) explicitly to that of a generator of the congruence subgroup $\Gamma_1(N) \subset \mathrm{SL}(2,\mathbb{Z})$, and we therefore refer to these as the \emph{modular coordinates}.

For the case in which $X$ is  elliptically fibered, we can perform a transformation composition by further twisting the branes with the line bundle $\mathcal{O}(-\tilde{S}_0)$ before and after the relative conifold transformation, where $[\tilde{S}_0] = [S_0] +\pi^\ast [\bar{K}_B]$, obtaining in turn the following monodromy transformation~\cite{Schimannek:2019ijf}
\begin{equation}
S: \tau \mapsto  -\frac{1}{\tau}\,, \quad z^A \mapsto \frac{z^A}{\tau}\,, \quad 
\hat{t}^\alpha \mapsto \hat{t}^\alpha- \frac{a^\alpha}{2}-\frac{1}{\tau}
b_{AB}^{\alpha}z^A z^B  + \mathcal{O}(e^{2\pi i \hat{t}^\alpha})
\,,
\label{eqn:Special}
\end{equation}
 We remark that in this transformation the induced shift on $\hat{t}^\alpha$ is determined by the anomaly coefficients introduced in~(\ref{eqn:AnCoef}). 

\noindent\textbf{Weyl group monodromies:} Another relevant monodromy for us is the one associated to a Weyl group symmetry. Such a type of monodromies occur when collapsing divisors to curves in $X$, giving rise to singularities in the stringy K\"ahler moduli space. For the cases in which $X$ has a torus fibration, the set of resolution divisors $E_{I_\kappa} \subset X$ are ruled surfaces over $\Sigma_\kappa \subset B$ whose $\mathbb{P}^1$-fibers are rational curves $C_{I_\kappa} \subset X$. 
Let $W_{I_\kappa}$ be the monodromies that encircle the singularities that occur when the curve $C_{I_\kappa}$ shrinks. 
Its action on the B-brane period associated to a vertical curve $C_f \subset X$, such that $[C_f]\in \mathrm{Mori}(X/B)$,  reads~\cite{Cota:2019cjx}
\begin{equation}
\label{eqn:WeylAction}
W_{I_\kappa}: \Pi_{C_f}\mapsto \Pi_{C_f} +(C_f \cdot E_{I_\kappa}) \Pi_{C_{I_\kappa}}\,.
\end{equation}

When $X$ admits an elliptic fibration, we can identify a set of curves $C^{I_\kappa} \subset X$ that are dual to the resolution divisors $E_{I_\kappa}$, and whose volume are $z^{I_\kappa}$, 
with the fundamental weights of $\mathfrak{g}_\kappa$. Such curves make a $\mathbb{Z}$-basis for the weight lattice $P(\mathfrak{g}_\kappa)$. 
Using~(\ref{eqn:WeylAction}), we then obtain the induced action on the K\"ahler moduli 
\begin{equation}
W_{I_\kappa}: z^{J_\nu} \mapsto z^{J_\nu}+ \delta^{\kappa\nu}\delta^{I J} C_{IK,\kappa} z^{K_\kappa}\,,
\end{equation}
where $C_{IJ,\kappa}$ is the Cartan matrix associated to $\mathfrak{g}_\kappa$. Therefore, we identify this monodromy with the  Weyl group action generated by reflections of negative simple roots of $\mathfrak{g}_\kappa$. 
Together, the monodromies $W_{I_\kappa}$ generate the action for the Weyl group $W(\mathfrak{g}_\kappa)$.

In the last section of this chapter, we use the monodromy subaction~(\ref{eqn:Usub}) to argue that an infinite tower of five-dimensional BPS states arises purely from geometrical data related to the counting of curves in the fiber direction. 
From a physics perspective, we identify such a set of geometrical invariants with Kaluza-Klein states that have a six-dimensional origin through F-theory compactifications, as discussed in section~\ref{sec:F-theory}.

\subsection{Kaluza-Klein tower of states via monodromies}

\label{sec:KKmodes}

To argue the presence of towers of BPS states associated to curve classes of $X/B$, 
let us recall a few facts about modularity. 
For starters, every elliptic curve $\mathbb{E}_\tau$ is analytically isomorphic to a complex torus $\mathbb{C}/\Lambda$, 
where 
\begin{equation}
\Lambda = \{n \omega_0 + m \omega_1 \mid n ,m \in \mathbb{Z} \} \subset \mathbb{C} \,,  
\label{eqn:Lattice}
\end{equation}
while $\omega_0$ and $ \omega_1$ are non-zero complex numbers. 
A pair  $(\omega_0, \omega_1)$ and $(\omega_0',\omega_1')$ generate the same lattice if and only if there is 
a transformation in $\mathrm{SL}(2,\mathbb{Z})$ relating them. 
As proved in~\cite{Toda:2011ma} and more recently discussed in~\cite{Bridgeland:2024ecw}, for smooth elliptic fibrations with at most $I_1$ Kodaira degenerations, a subset of BPS central charges associated with D2-D0 brane bound states can be identified with points of this lattice.  
All such D2-D0 bounded states carry identical BPS indices, giving rise to a tower of states in four and five dimensions.

More generally, let us now consider Calabi--Yau threefolds 
that arise as crepant resolutions of a singular elliptic or genus one fibrations.  
In these singular geometries, the torus fiber $\mathbb{E}_\tau$ 
degenerates over curves $\Sigma_\kappa\subset B$ or over isolated points in $B$. In the corresponding resolved geometries, the fiber over such loci decomposes into a finite union
of rational curves $C_i \subset X$, i.e.
\begin{equation}
\mathbb{E}_\tau = \bigcup_{i\in \mathcal{I}}  C_i\,.
\end{equation}
Here $\mathcal{I}$ is a finite index set associated with a given degeneration locus in $B$, in which either a codimension-one or codimension-two singularity occurs. Consequently, there is a finite family of rational curves $\{C_{i}\}$  
whose classes collectively span the relative Mori cone $\mathrm{Mori}(X/B)$. 

Similar as in~\cite{Bridgeland:2024ecw}, we consider the subcategory $D(\pi) \subset D^b(X)$ that is obtained by considering two-branes with support on 
curves in the fiber direction and the zero-brane with support on a point in $X$. 
Consquently, we obtain the following map
\begin{equation}
\mathrm{ch}: K(D(\pi)) \to M(\pi) \subset H^{2*}(X, \mathbb{Z})\,,
\end{equation}
where
\begin{equation}
M(\pi ) = \mathbb{Z} \gamma_0 \oplus_{i } \mathbb{Z} \gamma_i  \,.
\end{equation}
Here each $\gamma_i$ corresponds to the curve classes $[C_{i}]$ that are generators of $\mathrm{Mori}(X/B)$, while $\gamma_0$ is the cohomology class measuring the D0-brane charge. 
Let us note that there is a finite positive integral linear combination such that 
\begin{equation}
[\mathbb{E}_\tau] = \sum_{i  } k_i [C_i] \,,
\end{equation} 
where $[\mathbb{E}_\tau]$ is the curve class associated to the generic torus fiber $\mathbb{E}_\tau$ of $X$ and $k_i $ is a positive integer number. 
Accordingly, we can always find a two-dimensional sublattice $N(\pi) \subset M(\pi)$  associated to the generic fiber curve class, 
 given by
\begin{equation}
N(\pi) = \mathbb{Z} \gamma_0 \oplus \mathbb{Z}\left(\sum_{i} k_i \gamma_i \right) \,.
\end{equation}
Performing the composition of the central charge with the Chern character map, we obtain 
the period lattice 
\begin{equation}
\Pi_{\mathrm{B}}\left(n \gamma_0 + \sum_{i  } m_i \gamma_i\right) = n \Pi_{\mathrm{pt}} +\sum_{i } m_i \Pi_{C^i}  \,,
\label{eqn:PeriodLattice}
\end{equation}
Notice that $\Pi_{\mathrm{B}}(N(\pi))$ realizes a lattice isomorphic to $\Lambda$ in~(\ref{eqn:Lattice}), wheras the period lattice~(\ref{eqn:PeriodLattice}) is a subdivison of it. We want to consider the BPS indices associated with this lattice of central charges and their behaviour under transformations induced by auto-equivalences of $D(\pi)$. See~(\ref{eqn:BPSInd}) and Section~\ref{sec:HMS} for further discussion of BPS invariants.

Assuming that no wall-crossing occurs such as in the case of~\cite{Toda:2011ma}, taking the monodromy inverse $U^{-1}$ from~(\ref{eqn:Usub}), we notice that invariance of the D2-D0 BPS index under auto-equivalences 
of $D(\pi)$  
implies the symmetry $\Omega(\gamma + \sum_i k_i \gamma_i) = \Omega(\gamma )$ with $\gamma \in M(\pi)$.    
The latter assertion translates 
in terms of Gopakumar-Vafa invariants into the following periodicity relation
\begin{equation}
n_{0,[C]+ [\mathbb{E}_\tau]}^X = n_{0,[C]}^{X}\,,  \text{ for all } [C] \in \mathrm{Mori}(X/B) \,. 
\label{eqn:shiftGV}
\end{equation}
By F-theory/M-theory duality upon circle compactification, we interpret such a tower of five-dimensional BPS states---realized by the family of curve classes $[C] + n [\mathbb{E}_\tau]$ with $n \in \mathbb{N}$---as Kaluza-Klein modes derived  
from the masless spectrum of F-theory compactifications on $X$. 

To conclude this section, we summarize the main observations of~\cite{Paul-KonstantinOehlmann:2019jgr,Kashani-Poor:2019jyo}.  The multiplicities 
of massless matter in F-theory compactifications on torus fibrations $\pi: X \to B$ can be extracted from the enumerative geometry of fibral curves. 
Namely, there are
three different cases furnishing non-trivial genus zero Gopakumar--Vafa invariants for curve classes in $\mathrm{Mori}(X/B)$: 
\begin{enumerate}
\item  Isolated fibral curves occurring over codimension-two points $\Sigma_\kappa \cap \Sigma_\nu \subset B$, 
where $\Sigma_\kappa \subset B$ are codimension-one loci. 
M2-branes wrapping isolated fibral curves give rise to $n_{\mathbf{R}}$ charged hypermultiplets that transform under an irreducible representation $\mathbf{R}$
of $\mathfrak{g}_\kappa\oplus\mathfrak{g}_\nu$.  
The Gopakumar-Vafa invariant of an isolated curve $C$ is the number of
curves within the same class. Thus, we have that $n_{0,[C]}^{X} = n_{\mathbf{R}}$.
\item A fibral curve component of the generic fiber over a curve $\Sigma_\kappa \subset B$ of genus $g_\kappa$.
Such a curve gives $n_{0,[C]}^{X} = -\chi (\Sigma_\kappa) = 2g_\kappa -2$.  
\item A generic fiber curve, i.e. the torus fiber $C = \mathbb{E}_\tau$, yields $n_{0,[C]}^{X} = -\chi(X)$. 
\end{enumerate}

\noindent\textbf{Weyl group action:} Let us assume that $X$ admits an elliptic fibration. Consider a curve class $[C_f] \in \mathrm{Mori}(X/B)$. Given invariance of BPS indices under auto-quivalences of $D(\pi)$, we also obtain the following symmetry
\begin{equation}
n_{0,[C_f]}^{X} =n_{0,\mathcal{W}( [C_f])}^{X}\,. 
\label{eqn:WeylSymmetry}
\end{equation}
Here $\mathcal{W}$ is the Fourier-Mukai transform whose action is that of an element of the Weyl group $W(\mathfrak{g}_\kappa)$ of a given Lie algebra $\mathfrak{g}_\kappa$, such that it can be obtained through the generators described in~(\ref{eqn:WeylAction}).  For instance, for a given curve $C^{I_\kappa}$ that is dual to a resolution divisor $E_{I_\kappa}$, i.e.  a fundamental weight curve, all  associated weight curves in the $\mathfrak{g}_\kappa$-representation generated by the fundamental weight of $C^{I_\kappa}$~(\ref{eqn:Weights}) share the same D2-D0 BPS index. Similarly, all curves associated to negative roots of $\mathfrak{g}_\kappa$ are those appearing at codimension-one singularities and share the same D2-D0 BPS index among themselves. 

Having collected the necessary elements for monodromy transformations and enumerative geometry for the relative Mori cone $\mathrm{Mori}(X/B)$, we address in the upcoming chapter how the modularity of three-point couplings yields anomaly cancellation conditions.

\section{Supergravity anomaly equations via modularity}

Having established our working framework in topological string theory, we now describe the mechanism by which anomalies cancel via the BPS state counting associated with Kaluza--Klein towers. Quasi-Jacobi forms naturally encode both the anomalies and their cancellation through their modular transformation properties, which we review in Appendix~\ref{app:QJac}. The goal of this chapter is to derive anomaly cancellation conditions in F-theory, independently of the Green--Schwarz mechanism.

\label{sec:4}

\subsection{Modularity of three-point couplings on elliptic fibrations}

 Let consider smooth elliptically fibered Calabi--Yau threefolds $\pi : X \to B$. 
We want to study the automorphic form properties of the three-point couplings $C_{ijk}$ at base degree zero. 
By the latter object we mean the constant term in the Fourier-expansion of the form
\begin{equation}
C_{ijk}  = C_{ijk}^0 + \mathcal{O}(\widehat{Q}^\beta) \,,
\end{equation}
where $\widehat{Q}^\beta =\exp(2\pi i \hat{t}^\alpha \beta_\alpha)$ and $\beta = \beta_\alpha[ C^\alpha]$ with $C^\alpha$ being a basis of curves in $B$ whose classes span $\mathrm{Mori}(B)$,   
where we take the choice of modular coordinates~(\ref{eqn:KmodBasis}). We employ the action of~(\ref{eqn:Special}) on the corresponding K\"ahler moduli to obtain the modular transformation for $C^0_{ijk}$. At base degree zero, this transformation reads
\begin{equation}
S: (\tau, z^A,\hat{t}^\alpha) 
\mapsto
\left(\tau' =-\frac{1}{\tau},z^{A'} = \frac{z^A}{\tau}, \hat{t}^{\alpha'} = \hat{t}^\alpha -\frac{a^\alpha}{2} -\frac{1}{\tau} b_{AB}^\alpha z^A z^B \right)\,.
\end{equation} 
Here $a^\alpha$ and $b_{AB}^\alpha$ are the anomaly coefficients, introduced in~(\ref{eqn:AnCoef}). 
There are two different approaches for such a derivation that we outline next: 
\begin{enumerate}
\item The monodromy $S$ induces a tensorial  transformation rule on the three-point coupling with the following form\footnote{This transformation rule follows from~(\ref{eqn:3ptTrans}) and the factor $\tau^{-2}$ arises due to the monodromy action $S:\Pi_{\mathrm{pt}}\mapsto -\Pi_{\mathbb{E}_\tau}\,, \quad \Pi_{\mathbb{E}_\tau} \mapsto \Pi_{\mathrm{pt}}$. Moreover, in our basis of periods we relate $X^0 = -\Pi_{\mathrm{pt}}.$} 
\begin{equation}
    (C_{ijk}^0)'= \tau^{-2} \frac{\partial t^i}{\partial t^{i'}}\frac{\partial t^j}{\partial t^{j'}} \frac{\partial t^k}{\partial t^{k'}} C_{ijk}^0\,.
    \label{eqn:Const1}
\end{equation}
Here $t^{i'}$ denotes the transformed K\"ahler moduli $t^i$ under the action of $S$. By choosing the modular coordinates basis for the K\"ahler moduli, it is manifest that the induced transformation rule for $C^0_{ijk}$ resembles to modular transformations of quasi-Jacobi forms. 
\item Realizing the three-point coupling as the triple derivatives over the inhomogeneous holomorphic prepotential $F(\bm{t})$, we obtain the following transformation relation for its base degree zero contribution
\begin{equation}
C^0_{ijk}(\bm{t}') = c_{ijk} +  D_i D_j D_k  f \Big\vert_S \,.
\label{eqn:Const2}
\end{equation}
Here we introduced the notation $D_i :=\frac{1}{2\pi i} \partial_{t^i}$, where $f$ are the worldsheet instanton contributions of $F$ at base degree zero and $\vert_S$ denotes the action of $S$ on the corresponding automorphic form object. 
\end{enumerate}
We claim that matching both transformation rules~(\ref{eqn:Const1}) and~(\ref{eqn:Const2}) imposes algebraic  constraints that are equivalent to  anomaly cancellation conditions of both five- and six-dimensional supergravity theories.

In the following we discuss in more depth the necessary elements to derive the supergravity anomaly constraints via modularity of three-point couplings. 
First, the Jacobian matrix due to the monodromy $S$ acting on the modular coordinates  $(\tau, z^a,z^{I_\kappa},\hat{t}^\alpha)$ reads
\begin{align}
\frac{\partial t^i}{\partial t^{j'}} = 
\begin{pmatrix} 
\tau^2 & 0_b & 0_{J_\nu} & 0_\beta \\ 
\tau z^a & \delta^a_b \tau & 0^a_{J_\nu} & 0^a_\beta \\
\tau z^{I_\kappa} & 0^{I_\kappa}_b & \delta^{I_\kappa}_{J_\nu} \tau  & 0^{I_\kappa}_{\beta} \\
\frac{1}{2}b_{ab}^\alpha z^a z^b+\frac{1}{2}\delta_{\kappa \nu} C_{IJ,\kappa} b_{\kappa}^\alpha z^{I_\kappa} z^{J_\nu} & b_{ab}^\alpha z^a & \delta_{\kappa \nu} C_{IJ,\kappa} b_\kappa^\alpha z^{I_\kappa}z^{J_\nu} & \delta_{\beta}^\alpha
 \end{pmatrix} 
 + \mathcal{O}(\widehat{Q}^\beta)\,.
 \label{eqn:JacS}
\end{align}
Here, we expressed the base-degree-zero contributions to~(\ref{eqn:JacS}) in terms of the coordinates $t^i$ by first computing the Jacobian matrix $\partial_{t^j} t^{i'}$ and then inverting it. The remaining terms receive corrections that are exponentially suppressed in the base K\"ahler parameters. 

To illustrate how equation~(\ref{eqn:Const1}) works, let us consider as an example the base degree zero part for $C_{abc}$, 
which denotes the three-point coupling along the coordinates $z^a$, $z^b$ and $z^c$. 
We calculate~(\ref{eqn:Const1}) through the Jacobian matrix~(\ref{eqn:JacS}) and obtain
\begin{equation}
(C_{abc}^0)' =   \tau C_{abc}^0  -(b_{ab} \cdot b_{cd} +b_{ac} \cdot b_{bd} + b_{ad} \cdot b_{bc})z^d\,.
\label{eqn:CABC1}
\end{equation} 
Notice, on the one hand, the appearance of the left-hand-side of the pure abelian gauge anomaly equation~(\ref{eqn:pureGauge1}). On the other hand, it resembles the quasi-Jacobi form ring generator $A(\tau,z)$ that we discuss in Appendix~\ref{app:QJac} and transformas as
\begin{equation}
A(-1/\tau,z/\tau) = \tau A(\tau,z)+z\,.
\end{equation}
To make the connection between quasi-Jacobi forms and three-point couplings more precise, let us first discuss an essential ingredient in the anomaly cancellation conditions, namely the enumerative geometry of base degree zero curves on elliptically fibered Calabi--Yau threefolds. 

Before discussing in full generality the automorphic forms behind the base degree zero three-point couplings, for simplicity of the exposition, we first consider the case in which we encounter a gauge group $G = U(1)^r$ in an F-theory compactification, i.e. $X$ is an elliptic fibration with at most $I_2$ Kodaira singularities and a rank $r$ torsionless Mordell-Weil group. 
In such cases, the counting of Kaluza-Klein modes for the five-dimensional theory in section~\ref{sec:KKmodes}.  
is encapsulated by the following expression
\begin{equation}
f(\tau, \bm{z}) = -\frac{1}{2}\chi(X)\Phi^{0}(\tau,0)+  \sum_{\mathsf{q} \in \mathfrak{q}} n_{\mathsf{q}} \Phi^{\mathsf{q}}(\tau,\bm{z})  \,, 
\label{eqn:fibGV}
\end{equation}
where $\chi(X)$ is the Euler characteristic of $X$, $n_{\mathsf{q}}$ is the number of massless charged hypermultiplets with abelian gauge charge $\mathsf{q} = (\mathsf{q}_1,\ldots , \mathsf{q}_r)$ in the six-dimensional effective theory---alternatively the number of codimension-two singularities over a given class of isolated fibral curves~\cite{Kashani-Poor:2019jyo,Paul-KonstantinOehlmann:2019jgr},  
 $\mathfrak{q}$ is the set of charges in $\mathbb{Z}^r$ corresponding to all $U(1)^r$ massless charged hypermultiplets in the spectrum up to charge conjugation. 
Morever, we introduced the following modular object that 
has been identified in~\cite{Cota:2019cjx}
\begin{equation}
\Phi^{\mathsf{q}}(\tau,\bm{z}) =\mathrm{Li}_3(\bm{\zeta}^{\mathsf{q}})  + \sum_{n=1}^\infty\left( \mathrm{Li}_3(q^n \bm{\zeta}^{\mathsf{q}})  +\mathrm{Li}_3(q^n \bm{\zeta}^{-\mathsf{q}}) \right)\,.
\label{eqn:qPhi}
\end{equation}
Here $q = \exp(2\pi i \tau)$ and $\bm{\zeta}^{\mathsf{q}} = \exp(2\pi i \mathsf{q}_a z^a)$. Let us remark that the explicit expressions~(\ref{eqn:fibGV}) and~(\ref{eqn:qPhi}) can be derived from the fiber shift symmetry on Gopakumar-Vafa invariants~(\ref{eqn:shiftGV}). We outline next how this works.

The first term in~\eqref{eqn:qPhi} counts an isolated fibral curve $C_{\mathsf{q}}\subset X$ with charge $\mathsf{q}$ measured by
$
\mathsf{q}_a = C_{\mathsf{q}}\cdot \sigma(s_a).$
In the sum in~\eqref{eqn:qPhi}, we first encounter the set of curve classes $[C_{\mathsf{q}}] + n [\mathbb{E}_\tau]$ whose Gopakumar-Vafa invariants are related by~(\ref{eqn:shiftGV}).
Moreover, there is another contribution for $n=1$ in the sum that arises from counting the complementary splitting curve $C_{-\mathsf{q}}$, which is related to $C_{\mathsf{q}}$ by the class relation
$[C_{-\mathsf{q}}] = [\mathbb{E}_\tau] - [C_{\mathsf{q}}]$,
and therefore carries charge $-\mathsf{q}_a = C_{-\mathsf{q}}\cdot \sigma(s_a)$.
The remaining contributions of the  sum correspond the counting of curves $[C_{-\mathsf{q}}]+ n [\mathbb{E}_\tau]$, which are likewise related by~(\ref{eqn:shiftGV}).


So far we have established some working notation by using elliptically fibered Calabi--Yau threefolds leading to purely abelian gauge symmetries in the effective theory. We now generalize the form of  worldsheet instantons contributions for more general cases that also admit a non-abelian gauge group.

From now on, we assume that we have a general smooth elliptic fibration $\pi : X \to B$ whose associated gauge group in F-theory is $G$. The triple classical intersection relations in our K\"ahler form basis~(\ref{eqn:KmodBasis}) can be summarized within the prepotential classical contributions as follows~\cite{Park:2011ji,Weigand:2018rez}
\begin{align}
\begin{split}
F(t) \Big\vert_{\mathrm{class.}}=& \frac{1}{24}\bar{K}_B^2 \tau^3 +\frac{1}{2} \Omega_{\alpha \beta} \tau \hat{t}^{\alpha}\hat{t}^\beta  \\ 
&-\frac{1}{4}b_{ab} \cdot \bar{K}_B \tau z^a z^b  -\frac{1}{2}b_{ab,\alpha}
\hat{t}^\alpha z^a z^b+\frac{1}{6}c_{abc} z^a z^b z^c  \\
& -\frac{1}{4} \delta_{\kappa \lambda} \mathcal{C}_{IJ, \kappa} \left[\bar{K}_B \cdot b_{\kappa}\right] \tau z^{I_\kappa} z^{J_\lambda} 
-\frac{1}{2}  \delta_{\kappa \lambda }\mathcal{C}_{IJ,\kappa} b_{\kappa,\alpha} \hat{t}^\alpha z^{I_\kappa}
 z^{J_\lambda} \\ & +\frac{1}{6}
c_{I_\kappa J_\lambda K_{\mu}} z^{I_\kappa} z^{J_\lambda} z^{K_\mu} +\frac{1}{6} c_{a b I_{\kappa} } z^a z^b z^{I_\kappa}  +\frac{1}{6} c_{a I_{\kappa} J_{\lambda}} z^a z^{I_\kappa} z^{J_\kappa}\,,
\label{eqn:Fclass}
 \end{split}
\end{align}
where the explicit classical triple intersections can be obtained via the triple derivative relations $c_{ijk}= \partial_i \partial_j \partial_k F\vert_{\mathrm{class.}}$. 
On the other hand, the corresponding base degree zero worldsheet instanton contributions to the prepotential $f$ take the following form.
\\\\
\noindent\textbf{Claim-1:} For a smooth resolved elliptic fibration $\pi: X \to B$, we have the following form for $f$
\begin{empheq}[box=\fbox]{align}
 \begin{split}
 \label{eqn:Fq}
f(\tau, \bm{z}) = & -\frac{1}{2}\chi(X)\Phi^{0} (\tau,0)- \sum_{\kappa}\chi(\Sigma_\kappa) \Phi_{R^-_\kappa}(\tau,\bm{z}_\kappa)  \\
&
 +\sideset{}{'}\sum_{\mathbf{R},\mathsf{q}} n_{(\mathbf{R},\mathsf{q})} \Phi_{(\mathbf{R},\mathsf{q})}(\tau,\bm{z})
   \,.
 \end{split}
 \end{empheq}
Here $R^-_\kappa$ denotes the negative root system of $\mathfrak{g}_\kappa$ and we introduced the objects
\begin{equation}
    \Phi_{R_{\kappa}^-}(\tau,\bm{z}_\kappa) := \sum_{-\bm{\alpha}\in R_\kappa^-} \Phi^{-\bm{\alpha}}(\tau,\bm{z}_\kappa)\,, \quad 
\Phi_{(\mathbf{R},\mathsf{q})}(\tau,\bm{z}): = \sum_{\bm{\lambda} \in \Lambda (\mathbf{R)}} m_{\bm{\lambda}}\Phi^{(\bm{\lambda},\mathsf{q})}(\tau,\bm{z})\,,
     \label{eqn:PhiR}
  \end{equation}
  where $\mathbf{R}$ denotes an irreducible representation of the non-abelian part of the Lie algebra $\mathfrak{g}$ corresponding to $G$. 
  Moreover,  we express the negative roots in a Dynkin label basis,  $(\bm{\lambda},\mathsf{q}) = (\lambda_1, \ldots, \lambda_{\mathrm{dim}(\mathbf{R})},\mathsf{q}_1, \dots \mathsf{q}_r)$ and we use the same definition for $\Phi^{-\bm{\alpha}}$ and $\Phi^{(\bm{\lambda},\mathsf{q})}$ as in~(\ref{eqn:qPhi}).
  Furthermore, $\Lambda(\mathbf{R})$ is the set of weights contained in the convex hull of the Weyl orbit $W (\bm{\lambda}_+)$, where $\bm{\lambda_+}$ is the highest weight associated to $\mathbf{R}$ and $m_{\bm{\lambda}}$ 
  are the multiplicities for each weight $\bm{\lambda}$ for a given representation $\mathbf{R}$ that can be obtained via the Weyl characters~(\ref{eqn:WeylChar}). Finally, we indicate the prime sum over the last term in~(\ref{eqn:Fq}) to exclude the trivial and codimension-one singularities associated representations---already present in the first and second terms---and also to count modulo  charge conjugate representations $\bar{\mathbf{R}}$ and $-\mathsf{q}$ as we explain next. 

In definition~(\ref{eqn:PhiR}), similar to its abelian counter part, there are terms encoded in $\Phi^{(\bm{\lambda},\mathsf{q})}$ associated to splitting curves $C_{(\bm{\lambda},\mathsf{q})} \subset X$ such that their intersection with the resolution divisors $E_{i}$ and Shioda map $\sigma(s_a)$
determine the Dynkin labels $\lambda_{i} = (\bm{\lambda},\bm{\alpha}_i^\vee)$ and charge $\mathsf{q}_a$, respectively. In the same expression $\Phi^{(\bm{\lambda},\mathsf{q})}$, the complement curves $\mathbb{E}_\tau\backslash C_{(\bm{\lambda},\mathsf{q})}$ are encoded and their intersection with the resolution divisors $E_i$ and Shioda map $\sigma(s_a)$ gives the negative Dynkin label $-\lambda_i$ and $-\mathsf{q}_a$. Thus, $\Phi_{(\mathbf{R},\mathsf{q})}$ provides a counting for both charged matter representation that transforms under a given represntation $(\mathbf{R},\mathsf{q})$ as well as its conjugate pair in the conjugate representation $(\bar{\mathbf{R}},-\mathsf{q})$. 

We do not provide a strict proof for~(\ref{eqn:Fq}). Instead, we argue for its validity based on the symmetries~(\ref{eqn:shiftGV}) and~(\ref{eqn:WeylSymmetry}), the relationship of Gopakumar-Vafa invariants rational fibral and splitting curves with Euler characteristics of codimension-one and -two loci~\cite{Kashani-Poor:2019jyo,Paul-KonstantinOehlmann:2019jgr} and the   charge conjugation symmetry in counting states within a given representation. We also emphasize that we introduce the multiplicities $m_{\bm{\lambda}}$ as auxilary computational devices, since in most known cases one simply has $m_{\bm{\lambda}}=1$. However, we keep the expression conjectural as this may fail for exotic matter representations~\cite{Klevers:2017aku}. Having said this, from now on, to further simplify the notation
 we will rewrite~(\ref{eqn:Fq}) in compact form
 $f(\tau,\bm{z}) = \sum_{\mathcal{R}} n_{\mathcal{R}}\Phi_{\mathcal{R}}(\tau,\bm{z}) $, where $\mathcal{R} = (\mathbf{R},\mathsf{q})$\footnote{We will absorb $\Phi^0$ and $\Phi_{{R}_\kappa^-}$ as countings over the trivial and adjoint representations, respectively.}. We also denote the total weights of the non-abelian and abelian groups by $\bm{w} = (\bm{\lambda},\mathsf{q})$~\cite{Grimm:2013oga}, which lie in the lattice $\Lambda(\mathcal{R}) = \Lambda(\mathbf{R})\times \mathbb{Z}^r$.  

 The object $f$ can be realized as an Eichler integral as recently identified  in~\cite{Pioline:2025uov}. See also~\cite{Zagier2000} for further background and definitions. In contrast, our strategy here is to capture the modularity of the triple derivatives of $f$, i.e. the three-point couplings at base degree zero up to classical contributions. Note, however, that both descriptions encode the same enumerative geometry data.


\subsection{Quasi-Jacobi forms and three-point couplings}


First, we consider the base degree zero contributions to the three-point couplings $C_{\tau\tau\tau}$, $C_{\tau \tau A}$, $C_{\tau A B}$ and $C_{A BC}$.  
Starting from the fibral contributions of the genus zero prepotential~(\ref{eqn:fibGV}), we expect to obtain quasi-Jacobi forms of weight four, three, two and one respectively 
whenever taking the respective triple derivatives with respect to $\tau$ and $z^A$. 
It turns out that the modular objects that describe such type of three-point couplings are the deformed or twisted $n$-th Eisenstein series~\cite{2012arXiv1209.5628O}
 \begin{equation}
 J_n (\tau,z) = \delta_{n,1} \frac{\zeta}{\zeta -1} + B_n -n \sum_{k,l \geq 1} l^{n-1} (\zeta^k +(-1)^n \zeta^{-k} ) q^{kl} \,,
 \label{eqn:defEis}
 \end{equation}
 where $n \in \mathbb{Z}_{\geq 0}$, $B_n$ are the Bernoulli numbers, $z \in \mathbb{C}$ and $\zeta = \exp(2\pi i z)$. The name of $J_n$ is due to the properties
 $ J_{2k}(\tau,0) = B_{2k} E_{2k}(\tau)$ and $J_{2k-1}(\tau,0) = 0$ for $k\in \mathbb{N}$ with $E_{2k}(\tau)$ being the classical  Eisenstein series. 
Focusing at base degree zero generators of Gopakumar-Vafa invariants~(\ref{eqn:qPhi}), we then obtain the following identities.
 %
 
 \noindent\textbf{Claim-2:} 
 Consider a charge $\bm{w} \in \mathbb{Z}^m$ fibral Gromov-Witten potential $\Phi^{\bm{w}}(\tau,\bm{z})$, where $m\in \mathbb{N}$.  
Then its triple differential operators action relate to deformed Eisenstein series in the following way
\begin{empheq}[box=\fbox]{align}
\begin{split}
 D_\tau^3 \Phi^{\bm{w}}(\tau,\bm{z})  + \frac{1}{120} & =- \frac{1}{4} J_4(\tau, \bm{w} \cdot \bm{z}) \,,  \\
 D_\tau^2 D_{A} \Phi^{\bm{w}}(\tau,\bm{z}) & =-\frac{1}{3} w_A  J_3(\tau,\bm{w} \cdot \bm{z}) \,,\\
 D_\tau D_{A} D_{B} \Phi^{\bm{w}}(\tau,\bm{z})  -\frac{1}{12} w_A w_B & =-  \frac{1}{2} w_A w_B J_2(\tau,\bm{w} \cdot \bm{z}) \,,\\
 D_{A} D_{B} D_{C} \Phi^{\bm{w}}(\tau, \bm{z}) +\frac{1}{2}w_A w_B w_C & =-  w_A w_B w_C J_1(\tau,\bm{w} \cdot \bm{z}) \,.
 \end{split}
 \label{eqn:tripleDer}
\end{empheq}
Here we introduced the notation $\bm{w}\cdot\bm{z} = w_A z^A$, $D_\tau = \frac{1}{2\pi i} \partial_\tau$ and $D_A = \frac{1}{2\pi i}\partial_{z^A}$.

\textbf{Proof:} Consider the expansion
\begin{equation}
\Phi^{\bm{w}}(\tau,\bm{z}) = \sum_{k >0} \frac{\bm{\zeta}^{k\bm{w}}}{k^3} + \sum_{n,k>0} \frac{q^{k n}}{k^3} (\bm{\zeta}^{k \bm{w}} + \bm{\zeta}^{-k \bm{w}})\,.
\end{equation}
\begin{enumerate}
\item The first identity is given by
\begin{equation}
D_\tau^3 \Phi^{\bm{w}}(\tau,\bm{z}) =  \sum_{n,k>0} n^3 q^{k n} (\bm{\zeta}^{k \bm{w}} + \bm{\zeta}^{-k \bm{w}}) = -\frac{1}{4} \left( J_4(\tau,\bm{w} \cdot \bm{z})  - B_4\right) \,.
\end{equation}
\item The second identity is given by
\begin{align}
\begin{split}
D_\tau^2 D_{A} \Phi^{\bm{w}}(\tau,\bm{z}) &= w_A \sum_{n,k>0} n^2 q^{k n} (\bm{\zeta}^{k \bm{w}} - \bm{\zeta}^{-k \bm{w}})\\
&= -\frac{1}{3} w_A\left( J_3(\tau,\bm{w} \cdot \bm{z})  - B_3\right) \,.
\end{split}
\end{align}
\item The third identity is given by
\begin{align}
\begin{split}
D_\tau D_{A} D_{B} \Phi^{\bm{w}}(\tau,\bm{z}) &= w_A w_B \sum_{n,k>0} n q^{k n} (\bm{\zeta}^{k \bm{w}} + \bm{\zeta}^{-k \bm{w}}) \\
&= -\frac{1}{2}w_A w_B \left( J_2(\tau,\bm{w} \cdot \bm{z})  - B_2\right) \,.
\end{split}
\end{align}
\item The last identity gives a quasi-Jacobi form
\begin{align}
\begin{split}
D_{A} D_{B} D_{C} \Phi^{\bm{w}}(\tau,\bm{z}) & = w_A w_B w_C \left[\sum_{k>0} \bm{\zeta}^{k \mathsf{q}}+\sum_{n,k>0}  q^{k n} (\bm{\zeta}^{k \bm{w}} - \bm{\zeta}^{-k \bm{w}}) \right]   \\
& = -w_A w_B w_C \left( J_1(\tau, \bm{w} \cdot\bm{z}) -B_1 \right) \,.
\end{split}
\end{align}
Here we used the relation
\begin{equation}
\frac{\bm{\zeta}^{\bm{w}}}{1- \bm{\zeta}^{\bm{w}}} = \sum_{k>0} \bm{\zeta}^{k \bm{w}} \,,
\end{equation}
which is convergent and it follows from $\mathrm{Im}(z^A) >0$ since the latter quantities correspond to the real volume of curves $C^A \subset X$ that are objects in  $\mathrm{Mori}(X)$. 
\end{enumerate}

Using the triple derivative relations~(\ref{eqn:tripleDer}) together with the worldsheet instanton contributions~(\ref{eqn:Fq}), we can obtain now closed expressions for the base-degree zero three-point couplings. 
First, we obtain the \textit{pure gauge} base-degree zero three-point couplings:
\begin{align}
\begin{split}
     C_{ABC}^0 - c_{ABC} = -\sum_{\mathcal{R}}n_{\mathcal{R}} \sum_{\bm{w} \in \Lambda(\mathcal{R})}  m_{\bm{w}} 
     \left[ J_1(\tau,\bm{w} \cdot \bm{z}) -\frac{1}{2}\right]w_{A} w_{B} w_{C}  \,.
\end{split}
    \label{eqn:pureGauge}
\end{align}
Notice the dependence of the pure gauge three-point couplings on the first deformed Eisenstein series. Similarly, we introduce the \textit{mixed gauge-gravitational} three-point couplings:
\begin{align}
    \begin{split}
        C^0_{\tau A B} - c_{\tau AB} = -\frac{1}{2}\sum_{\mathcal{R}} n_{\mathcal{R}}\sum_{\bm{w} \in \Lambda(\mathcal{R})}  m_{\bm{w}} 
     \left[ J_2(\tau,\bm{w} \cdot \bm{z}) -\frac{1}{6}\right]w_{A} w_{B}
    \end{split}
    \label{eqn:mixedGG}
\end{align}
Notice here the presence of the second deformed Eisenstein series. In principle we could take other derivative combinations such as $C^0_{\tau \tau A}$. However, such a type of three-point couplings leads to trivial classical terms and hence will be not interesting to us.  Lastly we have the \textit{gravitational} three-point coupling
\begin{align}
    \begin{split}
        C_{\tau\tau\tau}^0 - \frac{1}{4}c_1(B)^2 = -\frac{1}{4} \sum_{\mathcal{R}}n_{\mathcal{R}}\sum_{\bm{w}\in\Lambda(\mathcal{R})} m_{\bm{w}}\left(J_4(\tau, \bm{w}\cdot \bm{z}) + \frac{1}{30}\right)\,.
    \end{split}
    \label{eqn:Grav}
\end{align}

Recall our claim stating that  the action~(\ref{eqn:Const1}) of the monodromy transformation $S$ on $C_{ijk}$ yields the anomaly counterterms appearing as modular transformation corrections, which are inherent to quasi-Jacobi forms. See for instance, Appendix~\ref{app:GSanomalies}. Now, by using directly the modular properties for the twisted or deformed Eiseinstein series~(\ref{eqn:nEisTrans})
on the explicit relations~(\ref{eqn:pureGauge}),~(\ref{eqn:mixedGG}) and~(\ref{eqn:Grav}), we obtain instead 
 one-loop anomaly polynomials in the anomalous transformation terms of quasi-Jacobi forms. Matching both transformed objects furnishes the anomaly cancellation conditions. To illustrate this phenomenon, we will consider the case for pure gauge bause degree zero three-point couplings in the following.

The transformation~(\ref{eqn:Const1}) for the pure abelian three-point coupling is of the form
\begin{equation}
(C_{ABC}^0)' =   \tau C_{ABC}^0+ \mathcal{A}_{ABCD}^{\mathrm{GS}}z^D\,,
\label{eqn:CABC1}
\end{equation} 
where $\mathcal{A}_{ABCD}^{\mathrm{GS}}$ are the ``anomalous" modular transformation coefficient terms whose precise expressions we include in Appendix~\ref{app:GSanomalies}.
Note that they correspond to Green--Schwarz counterterms appearing in the anomaly cancellation condition equations. 
Now, we perform the transformation  $S:(\tau,z^A)\mapsto (-1/\tau,z^A/\tau)$  on $C_{ABC}$ by using the expression~(\ref{eqn:pureGauge}). We obtain
\begin{align}
\begin{split}
(C_{ABC}^0)\Big\vert_{S} = & c_{ABC} - \frac{1}{2}\sum_{\mathcal{R}}\sum_{\bm{w}\in \Lambda(\mathcal{R})}n_{\mathcal{R}} m_{\bm{w}}w_A w_B
w_C\\
&- \sum_{\mathcal{R}}n_{\mathcal{R}} \sum_{\bm{w} \in \Lambda(\mathcal{R})}m_{\bm{w}} w_A w_B
w_C \left( w_D z^D+\tau  J_1(\tau,\bm{w}\cdot \bm{z}) \right)\,.
\label{eqn:CABC2}
\end{split}
\end{align}
Here we denote by $g(\tau,\bm{z})\vert_S := g(-1/\tau,\bm{z}/\tau)$ for some function $g$. Here we used the transformation relations~$J_1(-1/\tau,z/\tau)= z+\tau J_1(\tau,z)$ and that the constant terms transform trivially under $\mathrm{SL}(2,\mathbb{Z})$. 
Thus, imposing $(C_{ABC}^0)' \stackrel{!}{=} C_{ABC}^0\vert_S$, we conclude the following relations
\begin{align}
c_{ABC} = &\frac{1}{2}\sum_{\mathcal{R}}\sum_{\bm{w}\in \Lambda(\mathcal{R})}n_{\mathcal{R}} m_{\bm{w}}w_A w_B
w_C\,, \label{eqn:Grimm} \\
C_{ABC}^0 = &-\sum_{\mathcal{R}}n_{\mathcal{R}}\sum_{\bm{w}\in \Lambda(\mathcal{R})} m_{\bm{w}}w_A w_B
w_C J_1(\tau,\bm{w} \cdot \bm{z})
\label{eqn:qJ1}
\end{align}
and
\begin{equation}
\mathcal{A}_{ABCD}^{\mathrm{GS}} = \sum_{\mathcal{R}} n_{\mathcal{R}}\sum_{\bm{\mu}\in \Lambda(\mathcal{R})} m_{\bm{w}} w_A w_B w_C w_D\,. 
\label{eqn:gaugeAC}
\end{equation}
Let us remark that~(\ref{eqn:Grimm}) corresponds precisely to the induced pure gauge one-loop Chern-Simons coefficients equations in~\cite{Grimm:2013oga} obtained from the Kaluza-Klein reduction of F-theory on $X$. 
Finally, we emphasize that~(\ref{eqn:gaugeAC}) contains all pure gauge anomaly equations. We elaborate on this in what follows.\footnote{Interestingly, equation~(\ref{eqn:gaugeAC}) was also obtained in~\cite{Grimm:2015zea} by performing an elliptic transformation $z^A \mapsto z^A-\tau$ in which the elliptic parameters play the role of Wilson loop parameters in the Coulomb branch. }

Using equation~(\ref{eqn:gaugeAC}), together with the explicit coefficient terms $\mathcal{A}_{ABCD}^{\mathrm{GS}}$ in Appendix~\ref{app:GSanomalies} and the group theoretical identity~(\ref{eqn:traceId}), we obtain the alternative form for the pure gauge anomaly equations:~\cite[(B.19)]{Park:2011ji}
\begin{empheq}[box=\fbox]{align}
\begin{split}
b_{ab} \cdot b_{cd} + b_{ac} \cdot b_{bd} + b_{ad} \cdot b_{bc} &= \sideset{}{'}\sum_{\mathbf{R},\mathsf{q} } n_{(\mathbf{R},\mathsf{q})} \mathsf{q}_a \mathsf{q}_b \mathsf{q}_c \mathsf{q}_d\\
b_{\kappa}\cdot b_{\kappa}(\mathcal{C}_{IJ,\kappa} \mathcal{C}_{KL,\kappa}+\mathcal{C}_{IK,\kappa} 
\mathcal{C}_{JL,\kappa}+\mathcal{C}_{IL,\kappa}\mathcal{C}_{JK,\kappa}) &= \sideset{}{'}\sum_{\mathbf{R},\mathsf{q}} n_{(\mathbf{R},\mathsf{q})} \times \\
&\quad\quad\quad\quad\mathrm{tr}_{\mathbf{R}}\mathcal{T}_{I,\kappa}\mathcal{T}_{J,\kappa}\mathcal{T}_{K,\kappa}\mathcal{T}_{L,\kappa} \\
& - \frac{\chi(\Sigma_\kappa)}{2}\mathrm{tr}_{\mathbf{adj}_\kappa}\mathcal{T}_{I,\kappa}\mathcal{T}_{J,\kappa}\mathcal{T}_{K,\kappa}\mathcal{T}_{L,\kappa} \\
b_{\kappa} \cdot b_\nu \mathcal{C}_{IJ,\kappa} \mathcal{C}_{KL,\nu} &= \sideset{}{'}\sum_{\mathbf{R},\mathbf{S},\mathsf{q}}n_{\mathbf{(R\otimes\mathbf{S},\mathsf{q})}} \mathrm{tr}_{\mathbf{R}}\mathcal{T}_{I,\kappa} \mathcal{T}_{J,\kappa}\\
& \qquad \qquad \times\mathrm{tr}_{\mathbf{S}}\mathcal{T}_{K,\nu} \mathcal{T}_{L,\nu}\\
0 & = \sideset{}{'}\sum_{\mathbf{R},\mathsf{q}} n_{(\mathbf{R},\mathsf{q})} 
\mathsf{q}_a\mathrm{tr}_{\mathbf{R}}\mathcal{T}_{I,\kappa} \mathcal{T}_{J,\kappa} \mathcal{T}_{L,\kappa} \\
b_\kappa \cdot b_{ab} \mathcal{C}_{IJ,\kappa} & = \sideset{}{'}\sum_{\mathbf{R},\mathsf{q}} n_{(\mathbf{R},\mathsf{q})} \mathsf{q}_a \mathsf{q}_b\mathrm{tr}_{\mathbf{R}}\mathcal{T}_{I,\kappa}\mathcal{T}_{J,\kappa}
\end{split}
\end{empheq}
Considering contractions of the type $T_\kappa = t^I \mathcal{T}_{I,\kappa}$ and $T_\nu = s^I \mathcal{T}_{I,\nu}$, together with the group theory coefficients $A_{\mathbf{R}}$, $B_{\mathbf{R}}$, $C_{\mathbf{R}}$ and $E_{\mathbf{R}}$ relations~(\ref{eqn:groupCoef})
we obtain the pure gauge anomaly equations of~(\ref{eqn:pureGauge1}). 

The same method can be used to derive the mixed- and pure gravitational anomaly equations from the modular transformations associated to the mixed- and pure gravitational three-point couplings respectively. However, we will use a simpler method in the following section to derive the remaining anomaly cancellation conditions that connect with the holomorphic anomaly equations appearing in topological string theory~\cite{Bershadsky:1993cx}.  

 \subsection{Anomalies cancellation from holomorphic anomaly equations}

We now describe the supergravity anomalies cancellation mechanism in terms of the so-called modular anomaly equations.  We derive them from the holomorphic anomaly equations of four-point couplings of topological string theory. 

In the two-dimensional worldsheet description of the B-model topological string theory, the four-point functions are described as correlation functions among three chiral primary fields,  together with an additional insertion of a descendant field. Geometrically, such an extra insertion is described by applying a further derivative
 \begin{equation}
     C_{i_1 i_2 i_3 i_4} = \mathcal{D}_{i_4} C_{i_1 i_2 i_3}  \subset  \mathrm{Sym}^4(T_{1,0}^\ast(\mathcal{M}_{\mathrm{cs}}))\otimes \mathcal{L}^2\,,
 \end{equation}
 where $\mathcal{D}_i$ is a covariant derivative that becomes flat in the large volume limit. 
 Such four-point correlator functions satisfy the holomorphic anomaly equation~\cite{Bershadsky:1993cx,Lee:2020blx}
 \begin{equation}
 \bar{\partial}_{\bar{\imath}} C_{i_1 i_2 i_3 i_4}  = \frac{1}{8} \bar{C}_{\bar{\imath}}^{jk} \sum_{\sigma \in S_4} C_{j i_{\sigma(1)} i_{\sigma(2)}}  C_{k i_{\sigma(3)} i_{\sigma(4)}} 
  -\sum_{s = 1}^4 G_{\bar{\imath} i_s}  C_{i_1 \cdots i_{s-1} i_{s+1} \cdots i_4} \,, 
  \label{eqn:BCOV4pt}
 \end{equation} 
 where $\bar{C}_{\bar{\imath}}^{jk} = \bar{C}_{\bar{\imath}\bar{\jmath}\bar{k}} e^{2K} G^{j \bar{\jmath}} G^{k \bar{k}}$, $G_{\bar{\imath}j} = \partial_{\bar{\imath}}\partial_j K$ is the Zamolodchikov metric on the moduli space $\mathcal{M}_{\mathrm{cs}}$ and $K$ is its associated K\"ahler potential.
 In the context that we consider the A-model on a smooth elliptically fibered Calabi--Yau threefold $\pi: X \to B $, we argue how the holomorphic anomaly equation~(\ref{eqn:BCOV4pt}) leads to the next modular anomaly equation---in the topological limit $\bar{t}_{\bar{\imath}}\to  \infty$
 \begin{empheq}[box=\fbox]{align}
 \begin{split}
 \frac{\partial}{\partial E_2} C_{ABCD} = & \frac{1}{96} \Omega^{\alpha \beta} \sum_{\sigma \in S_4} C_{\alpha \sigma(A) \sigma(B)} C_{\beta \sigma(C) \sigma(D)}
 ,\\
 \frac{\partial}{\partial E_2}C_{\tau \tau AB} = & \frac{1}{6} C_{\tau A B} \,, \\ 
 \frac{\partial}{\partial E_2} C_{\tau \tau \tau \tau} = &\frac{1}{3} C_{\tau \tau \tau}\,. 
 \end{split}
 \label{eqn:E2Grav}
 \end{empheq}
 Recall that $\Omega^{\alpha \beta} = [C^\alpha ]\cdot [C^\beta]$,  where $[C^\alpha]$ is a basis of of curve classes in $B$. Moreover, the index $A$ denotes any K\"ahler parameter $z^{A} \in \{z^a, z^{i_I} \}$. For later convenience, let us define $C_{ijkl}^{\mathrm{GS}}$ to be the right-hand side of equations~(\ref{eqn:E2Grav}). 
 As we will see, equations~(\ref{eqn:E2Grav}) encode---respectively---the pure gravitational, mixed-gravitational, and pure gravitational anomaly constraints.   
 
To derive the modular anomaly equations~(\ref{eqn:E2Grav}), we take the the small fiber limit approach
by considering the regime in which the base moduli $\hat{t}^\alpha$ scale as $\hat{t}^\alpha /h $ with $h\to 0$
while keeping the fiber parameters $\tau$ and $z^{A}$ fixed~\cite{Cota:2019cjx, Lee:2020blx}. 
In other words, we are interested in the limit
\begin{equation}
\mathrm{Im}(\hat{t}^\alpha) \to \infty \quad \text{such that} \quad  \frac{\mathrm{Im}(\tau)}{\mathrm{Im}(\hat{t}^\alpha)} \to 0 \,, \quad \frac{\mathrm{Im}(z^A)}{\mathrm{Im}(\hat{t}^\alpha)} \to 0 \,,\quad \frac{\mathrm{Im}(z^{A})}{\mathrm{Im}(\tau)} = \mathcal{O}(1) \,.  
\label{eqn:smallFib}
\end{equation}  
In particular, we want to consider the anti-holomorphic derivative $\bar{\partial}_{\bar{\tau}}$ for the BCOV equation~(\ref{eqn:BCOV4pt}). 
Thus, we compute the required objects on the right-hand side of the BCOV equation in the small fiber regime 
and obtain the leading order contributions
\begin{equation}
C_{\bar{\tau}}^{jk} = 
\left( \begin{array}{c|c}
  C_{\bar{\tau}}^{\hat{A} \hat{B}} &  C_{\bar{\tau}}^{\hat{A} \beta} \\
   \midrule
   C_{\bar{\tau}}^{\alpha \hat{B}} & C_{\bar{\tau}}^{\alpha \beta} \\
\end{array}\right) \, \quad \stackrel{(\ref{eqn:smallFib})}{\Rightarrow} \quad
C_{\bar{\tau}}^{jk} = 
\left( \begin{array}{c|c}
   \mathcal{O}(h) & \mathcal{O}(h) \\
   \midrule
   \mathcal{O}(h) & \frac{1}{(2\pi)^2 4\tau_2} \Omega^{\alpha \beta} + \mathcal{O}(h)  \\
\end{array}\right) \,.
\end{equation}
Here we use the label $\hat{A}\in \{\tau, A\}$. 
Moreover, we obtain\footnote{Here we use the normalization $G_{\bar{\imath}j} = \frac{-\bar{\partial}_{\bar{\imath}}}{2\pi i} \frac{\partial_j}{2\pi i} K$.}
\begin{equation}
G_{\bar{\imath}j} \stackrel{(\ref{eqn:smallFib})}{=} \begin{cases} \frac{1}{16 \pi^2 \tau_2^2} + \mathcal{O}(h) &\text{ if } \bar{\imath} = \bar{\tau} \text{ and } j = \tau\,, \\
 \mathcal{O}(h)&\text{ otherwise }\,.\end{cases} \,.
\end{equation}
Putting all pieces together in~(\ref{eqn:BCOV4pt}) we obtain for the cases relevant to us
\begin{align}
\begin{split}
\frac{\tau_2^2}{3} \frac{4\pi^2 }{2\pi i} \bar{\partial}_{\bar{\tau}} &C_{ABCD} =\frac{1}{96} \Omega^{\alpha \beta} \sum_{\sigma \in S_4} C_{\alpha \sigma(A) \sigma(B)} C_{\beta \sigma(C) \sigma(D)} + \mathcal{O}(h)\,, \\
\frac{\tau_2^2}{3} \frac{4\pi^2 }{2\pi i} \bar{\partial}_{\bar{\tau}} & C_{\tau\tau A B} =\frac{1}{6} C_{\tau A B} +\mathcal{O}(h) \,, \quad \frac{\tau_2^2}{3} \frac{4\pi^2 }{2\pi i} \bar{\partial}_{\bar{\tau}} C_{\tau\tau\tau\tau}=\frac{1}{3} C_{\tau\tau\tau} + \mathcal{O}(h)\,.
\end{split}
\end{align}
By taking the exchange 
\begin{equation}
-\frac{ 2 \pi i}{3}  \tau_2^2\bar{\partial}_{\bar{\tau}} \leftrightarrow \partial_{E_2}
\end{equation}
in the holomorphic limit $\bar{t}_{\bar{\imath}} \to \infty$~\cite{Cota:2019cjx, Lee:2020blx} 
and then the limit $h\to 0$, we obtain
~(\ref{eqn:E2Grav}). See more details about the anomaly operator $\partial_{E_2}$ in Appendix~\ref{app:QJac}, (\ref{eqn:delE2}). We also identify here the holomorphic limit with the constant term map~(\ref{eqn:ctMap}).  
 
Finally, we argue how the anomaly operator $\partial_{E_2}$ acting on the four-point coupling measures the one-loop anomaly coefficient for the $\mathcal{N}=(1,0)$ six-dimensional supergravity theories arising from F-theory on elliptically fibered Calabi--Yau threefolds. 
 To this end, we first need to obtain the quasi-Jacobi form expression associated to the fourth elliptic derivatives over $\Phi^{\bm{w}}$
 \begin{equation}
 D_{A} D_{B} D_{C} D_{D} \Phi^{\bm{w}}(\tau,\bm{z})
 = w_{A} w_{B} w_{C} w_{D} \left[ - \wp(\tau , \bm{w} \cdot \bm{z}) +\frac{1}{12}E_2(\tau)\right]\,,
 \end{equation}
where $\wp(\tau,z)$ is the Weierstra\ss~elliptic function, defined in~(\ref{eqn:WP}). The last expression is obtained easily from  
the algebraic relations in~(\ref{eqn:DerGens}). 
Thus, we obtain the following expression
\begin{align}
\begin{split}
\frac{\partial}{\partial E_2} C_{ABCD} \Big \vert_{\beta = 0} =& \frac{\partial}{\partial E_2} D_{A} D_{B} D_{C} D_{D} f(\tau,\bm{z}) = \frac{1}{12} \sum_{\mathcal{R}}  n_{\mathcal{R}} \sum_{\bm{w}\in \Lambda(\mathcal{R})} m_{\bm{w}}w_{A} w_{B} w_{C} w_{D} \\
= & \frac{1}{12} \sum_{\mathcal{R}} n_{\mathcal{R}} \mathrm{tr}_{\mathcal{R}} \mathcal{T}_{A} \mathcal{T}_{B} \mathcal{T}_{C} \mathcal{T}_{D} \equiv  2 \mathcal{A}_{ABCD}^{(6)}\,.
\end{split}
\label{eqn:ABCD}
\end{align}
Here $\beta \in H_2(B,\mathbb{Z})$ and we use $\cdot\vert_{\beta =0}$ to denote base degree zero contribution, while $\{\mathcal{T}_{A}\}$ is the coroot basis for the Cartan subalgebra of $\mathfrak{g}$. 
Notice that here we identified the action of the modular anomaly operator on the base degree zero contribution to the four-point coupling $C_{ABCD}$ with a corresponding pure gauge six-dimensional one-loop anomaly polynomial $\mathcal{A}_{ABCD}^{(6)}$. 
Similarly, we encounter mixed- and pure gravitational anomaly polynomials $\mathcal{A}_{AB}^{(6)}$ and $\mathcal{A}_{\mathrm{grav}}^{(6)}$ as follows
\begin{align}
\begin{split}
   \frac{\partial}{\partial E_2} C_{\tau\tau A B} \Big\vert_{\beta=0} \Big\vert_{q^0} 
  & = \frac{1}{72}\sum_{\mathcal{R}} n_{\mathcal{R}} \mathrm{Tr}_{\mathcal{R}}\mathcal{T}_A\mathcal{T}_B\equiv \mathcal{A}^{(6)}_{AB}\,, \\
     \frac{\partial}{\partial E_2} C_{\tau\tau \tau\tau} \Big\vert_{\beta=0} \Big\vert_{q^0} & 
     = \frac{1}{360} \sum_{\mathcal{R}}n_{\mathcal{R}}\mathrm{dim}(\mathcal{R}) \equiv \mathcal{A}_{\mathrm{grav}}^{(6)}\,.
     \end{split}
     \label{eqn:Amix}
\end{align}
Here $q^0$ denotes the zeroth coefficient over the Fourier expansion with respect to the $q = \exp(2\pi i\tau)$ parameter. Thus, we observe that the left-hand-side of the modular anomaly equations~(\ref{eqn:E2Grav}) encode the six-dimensional one-loop anomaly polynomials up to scaling factors. 

Let us denote the right-hand side of equation~(\ref{eqn:E2Grav}) by $C_{ijkl}^{\mathrm{GS}}$. We note that such an object can only yield classical term contributions for the pure gauge anomaly case, which is due to its form given by products of three-point couplings of the form $C_{\alpha AB}$ at base degree zero. A straightforward calculation leads us to obtain 
\begin{equation}
C_{ABCD}^{\mathrm{GS}}=\frac{1}{96} \Omega^{\alpha \beta} \sum_{\sigma \in S_4} C_{\alpha \sigma(A) \sigma(B)} C_{\beta \sigma(C) \sigma(D)} =  2\mathcal{A}_{ABCD}^{\mathrm{GS}}\,,
\end{equation}
whose explicit expressions for $\mathcal{A}_{ABCD}^{\mathrm{GS}}$ we reported in Appendix~\ref{app:GSanomalies}, as explained in previous section. However, for the mixed- and pure-gravitational cases at first glance it seems to provide trivial equations due to the quasi-Jacobi form relations
\begin{equation}
    \frac{\partial}{\partial{E_2}}D_\tau J_2(\tau,z) = \frac{1}{6}J_2(\tau,z)  \,, \quad \frac{\partial}{\partial{E_2}}D_\tau J_4(\tau,z) = \frac{1}{3}J_4(\tau,z)\,.
\end{equation}
obtained from~(\ref{eqn:idJ2J4}). However, the critical mechanism occurs when matching the zeroth coefficient $q^0$ with the classical intersection data, i.e.
\begin{equation}
    C_{\tau\tau AB}^{\mathrm{GS}} \Big\vert_{q^0} = 
    \frac{1}{6}c_{\tau AB}\,, \quad   C_{\tau\tau \tau\tau}^{\mathrm{GS}} \Big\vert_{q^0} = 
    \frac{1}{12}\bar{K}_B^2\,.
    \label{eqn:AmixGS}
\end{equation}
Thus, matching~(\ref{eqn:Amix}) with~(\ref{eqn:AmixGS}) we obtain the alternative form for the mixed gauge gravitational anomaly equations~\cite[(B.19)]{Park:2011ji}
\begin{empheq}[box=\fbox]{align}
\begin{split}
\bar{K}_B\cdot b_{ab} & = \frac{1}{6} \sum_{\mathbf{R},\mathsf{q}} n_{(\mathbf{R},\mathsf{q})} \mathsf{q}_a \mathsf{q}_b \\
\bar{K}_B \cdot b_\kappa \mathcal{C}_{IJ,\kappa} & = \frac{1}{6}\left(-\frac{1}{2}\chi(\Sigma_\kappa)\mathrm{tr}_{\mathbf{adj}_\kappa}\mathcal{T}_{I,\kappa}\mathcal{T}_{J,\kappa}+\sum_{\mathbf{R},\mathsf{q}}n_{(\mathbf{R},\mathsf{q})} \mathrm{tr}_{\mathbf{R}} \mathcal{T}_{I,\kappa} \mathcal{T}_{J,\kappa}\right)\,,
\label{eqn:mixRaw}
    \end{split}
\end{empheq}
which reduce to the familar expression for mixed gravitational anomaly equations in~(\ref{eqn:mixedAn1}) by taking contractions of the form $T_\kappa = t^I \mathcal{T}_{I,\kappa}$ and using the identities~(\ref{eqn:groupCoef}).  
Morever, from the pure gravitational case, we obtain
\begin{empheq}[box=\fbox]{equation}
    \frac{1}{2} \chi(X) + 30 \bar{K}_B^2 = \sideset{}{'}\sum_{\mathbf{R},\mathsf{q}}n_{(\mathbf{R},\mathsf{q})}\mathrm{dim}(\mathbf{R}) -\frac{1}{2}\sum_\kappa\chi(\Sigma_\kappa) [\mathrm{dim}(\mathbf{adj}_\kappa) - \mathrm{rk}(\mathfrak{g}_\kappa)]\,,
    \label{eqn:gravGeom}
\end{empheq}
which is precisely the pure gravitational equation stated in Theorem 3.1 in~\cite{Grassi:2000we}, also appearing in~\cite[(3.8)]{Park:2011ji}.

 \section{Examples}

 We present a couple of explicit examples to clarify the general discussion.
 \label{sec:eg}
\subsection{$G=U(1)$ examples}
Let us consider a class of elliptic fibrations with a single rational section that are Morrison-Park models with base $B = \mathbb{P}^2$. 
 See~\cite[Section 7.2]{Weigand:2018rez}.
 Then we have a family of three K\"ahler parameter Calabi-Yau threefolds whose Euler characteristic reads
 \begin{equation}
 \chi = -2 \left(4 m^2-3 m n-12 m+2 n^2-12 n+108\right)
 \end{equation}
 for the given choice of pair of integers
 \begin{equation}
 (m,n) \in \{(0,-3), (1,-2), (2,-1), (3,0), (4,1), (5,2), (6,3) \}\,.
 \label{eq:int}
 \end{equation}
 We have the following number of masless charged hypermultiplets
 \begin{equation}
 n_{\mathsf{q}=1} = -3 m^2-2 m n+39 m+n^2-15 n+54 \,, \quad n_{\mathsf{q}=2} = 2 m^2-m n\,. 
 \end{equation}
 Moreover, the height-pairing reads~\cite{Klevers:2014bqa}
 \begin{equation}
 b = 6 H + 2 H m \,,
 \end{equation}
 where $H$ is the hyperplane class in $\mathbb{P}^2$. 
 Given that $\bar{K}_B = 3H$, a simple calculation shows that the following constraints hold 
 for all models in the family of Morrison-Park models determined by the integers~(\ref{eq:int})
 \begin{equation}
 60 \bar{K}_B^2 = -\chi + 2 n_{\mathsf{q}=1}+2 n_{\mathsf{q}=2}\,, \quad b \cdot \bar{K}_B = \frac{1}{6} \sum_{\mathsf{q} \in \{1,2\} } n_{\mathsf{q}} \mathsf{q}^2 \,,  \quad 3 b^2 = \sum_{\mathsf{q} \in \{1,2\} } n_{\mathsf{q}} \mathsf{q}^4 \,.
 \end{equation}
 Notice that the case $(m,n) = (3,0)$ has been discussed in section 7.2.2 in~\cite{Cota:2019cjx}\,.

 \subsection{$G = SU(3)$ examples}
 We consider now a class of  ellipitic fibrations $\pi : X_{\mathfrak{a}_2} \to \mathbb{P}^2$, whose corresponding F-theory realizations leads to a gauge group $\mathrm{SU}(3)$. Using the methods of~\cite{Klevers:2014bqa}, we construct a geometry that contains rational curves over $I_3$ singularities as well as isolated fibral curves associated to codimension-two singulariteis, giving rise to massless matter that transforms in the fundamental representation $\mathbf{3}$. 

 These geometries can be specified by the following toric data via Batyrev's construction of Calabi--Yau hypersurfaces in toric varieties~\cite{Batyrev:1994pg}. Here we only list the vertices $\nu^*$ of a reflexive polytope which defines a toric ambient space in which $X_{\mathfrak{a}_2}$ is a hypersurface. The data reads
 \begin{align}
\begin{blockarray}{crrrrl}
	&&&&&\\
\begin{block}{c(rrrr)l}
         \nu_1^*&  0 &  0 &  -2 & -3  \\ 
         \nu_2^*&  0&  0 &  -1 & -2  \\ 
         \nu_3^*& 0&  0 &  0 & -1   \\
	\nu_4^*& 0&  0 &  1&  0   \\
	\nu_5^*& 0 & 0& 0 & 1    \\
        \nu_6^*&   1&   0& m-3  & 2m-3  \\
          \nu_7^*& 0&  1 & 0 & 0  \\
          \nu_8^*& -1 & -1& 0 & 0 &  \\
\end{block}
\end{blockarray}\,.
\label{eqn:ellipticFibOverBlF2toricdata}
\end{align}
Here we consider the values $m \in\{-2,-1,0,1,2,3\}$.
 The Euler characteristic of these geometries is given by
 \begin{equation}
\chi=-216+72m-12m^2\,.
 \end{equation}
 Using the modular coordinates parametrization $\omega = \hat{S}_0 \tau + E_1 z^1 +E_2 z^2 +  \pi^\ast H \hat{t}$,
 the intersection ring of $X_{\mathfrak{a}_2}$ can be brought into the following form 
 \begin{equation}
     F\vert_{\mathrm{class.}} = \frac{\bar{K}_B^2}{24}\tau^3  +\frac{1}{2} \tau \hat{t}^2- \frac{1}{4}( b_\kappa \cdot \bar{K}_B) \mathcal{C}_{IJ,\kappa}z^I z^J-\frac{1}{2}\mathcal{C}_{IJ,\kappa}\hat{t} z^I z^J +\frac{1}{6}c_{IJK}z^I z^J z^K\,.
 \end{equation}
 Here $H$ is the hyperplance class of $\mathbb{P}^2$ and $\mathcal{C}_{IJ,\kappa}$ is the  coroot matrix associated to $\mathfrak{a}_2$. 
We also identify the codimension-one singular curve class $b_\kappa = [\Sigma_\kappa]$ and its genus as follows
 \begin{equation}
     b_\kappa = (3+m)H\,, \quad g(\Sigma_\kappa) = \frac{1}{2}(3m+m^2) +1\,.
 \end{equation}
 We summarize next the relevant curve classes for the Gromov--Witten theory of $X_{\mathfrak{a}_2}$ at base degree zero. 
 
 Using standard mirror symmetry computations~\cite{Hosono:1993qy,Hosono:1994ax}, we compute the genus zero Gopakumar--Vafa invariants and obtain that the relative Mori cone is spanned as follows 
 \begin{equation}
 \mathrm{Mori}(X/B) = \langle [C_{-\bm{\alpha}_1}], [C_{-\bm{\alpha}_2}], [C_{\bm{\omega}_1}]\rangle\,,
 \end{equation}
 where $C_{-\bm{\alpha}_I}$ are curves whose intersections with the resolution divisors $E_{I}\cdot C_{-\alpha_J}$ reproduces the negative Cartan matrix of $\mathfrak{a}_2$,  while $C_{\bm{\omega}_I}$ that is dual to $E_{J}$.
 We furthermore encounter the following curve classes with non-zero genus zero Gopakumar-Vafa invariant
 \begin{equation}
  \mathbf{3}:    [C_{\bm{\omega}_1}]\,, \quad[C_{\bm{\omega}_1-\bm{\alpha}_1}] =[C_{\bm{\omega}_1}]+ [C_{-\bm{\alpha}_1}] \,, \quad   [C_{\bm{\omega}_1-\bm{\alpha}_1-\bm{\alpha}_2}] = [C_{\bm{\omega}_1-\bm{\alpha}_1}] +[C_{-\bm{\alpha}_2}] \,.
 \end{equation}
 The label $\mathbf{3}$ indicates that the intersections of these curves with $E_{I}$ yield the Dynkin labels of the weights in the fundamental representation $\mathbf{3}$. As expected from~(\ref{eqn:WeylAction}), we obtain
 \begin{equation}
     n_{\mathbf{3}} = n^X_{0,[C_{\bm{\omega_1}}]} = n^X_{0,[C_{\bm{\omega}_1-\bm{\alpha}_1}]}=n^X_{0,[C_{\bm{\omega}_1-\bm{\alpha}_1-\bm{\alpha}_2}]} = 54 + 9 m - 3 m^2\,.
 \end{equation}
 Similarly, we find
 \begin{equation}
     2g(\Sigma_\kappa)-2 = n_{0,[C_{-\bm{\alpha}_1}]}^X= n_{0,[C_{-\bm{\alpha}_2}]}^X =  n_{0,[C_{-\bm{\alpha}_1-\bm{\alpha_2}}]}^X\,.
 \end{equation}
  Moreover, we note the elliptic fiber class satisfies the relation 
  \begin{equation}
  [\mathbb{E}_\tau] = [C_{\bm{\omega_1}}]+ [C_{\bm{\omega_1-\bm{\alpha}_1}}]+[C_{\bm{\omega_1-\bm{\alpha}_1-\bm{\alpha}_2}}]= [C_{-\bm{\alpha_1}}]+ [C_{-\bm{\alpha}_2}]+[C_{\bm{-\bm{\alpha}_1-\bm{\alpha}_2}}]
  \end{equation}
  and its genus zero Gopakumar--Vafa invariant is $n_{0,[\mathbb{E}_\tau]}^X = - \chi$. In addition, there is a set of complementary curve classes, which we define by
  \begin{equation}
      \bar{\mathbf{3}}: [C_{\bm{\omega}_2}]\,, \quad [C_{\bm{\omega}_2-\bm{\alpha}_2}]=[C_{\bm{\omega}_2}]+[C_{-\bm{\alpha}_2}]\,, \quad [C_{\bm{\omega}_2-\bm{\alpha}_1-\bm{\alpha_2}}] = [C_{\bm{\omega_2}-\bm{\alpha}_2}] + [C_{-\bm{\alpha_1}}]\,,
  \end{equation}
  where $[C_{\bm{\omega}_2}]= [\mathbb{E}_\tau]- [C_{\bm{\omega_1}-\bm{\alpha_1}-\bm{\alpha}_2}]$. Thhe label $\bar{\mathbf{3}}$ indicates that the intersections of these curves with $E_{I}$ yield the Dynkin labels of the weights in the anti-fundamental representation $\bar{\mathbf{3}}$. Similarly, we  define 
  \begin{equation}
     [ C_{\bm{\alpha_1}}] = [\mathbb{E}_\tau ] - [C_{-\bm{\alpha_1}}] \,, \text{ } [C_{\bm{\alpha}_2}] = [\mathbb{E}_\tau]- [C_{-\bm{\alpha}_2}]\,, \text{ } [C_{\bm{\alpha}_1 +\bm{\alpha}_2}] = [\mathbb{E}_\tau] - [C_{-\bm{\alpha}_1 -\bm{\alpha}_2}]\,.
  \end{equation}
  We then obtain the relations 
  \begin{equation}
           n_{\bar{\mathbf{3}}} = n^X_{0,[C_{\bm{\omega_2}}]} = n^X_{0,[C_{\bm{\omega}_2-\bm{\alpha}_2}]}=n^X_{0,[C_{\bm{\omega}_2-\bm{\alpha}_1-\bm{\alpha}_2}]} =n_{\mathbf{3}}\,
  \end{equation}
  and
  \begin{equation}
   2g(\Sigma_\kappa)-2 = n_{0,[C_{\bm{\alpha}_1}]}^X= n_{0,[C_{\bm{\alpha}_2}]}^X =  n_{0,[C_{\bm{\alpha}_1+\bm{\alpha_2}}]}^X\,.
   \end{equation}
   The remaining genus zero Gopakumar--Vafa invariants of $X_{\mathfrak{a}_2}$ at base degree zero follow from~(\ref{eqn:shiftGV}) together with the curve classes described above.  
 
In brief, the worldsheet instanton corrections to the prepotential of $X_{\mathfrak{a}_2}$ at base degree zero take the form
 \begin{equation}
     f(\tau,\bm{z}) =-\frac{1}{2} \chi \Phi^0(\tau,\bm{z}) +[2g(\Sigma_\kappa)-2]\Phi_{R^-}(\tau,\bm{z})+ n_{\mathbf{3}} \Phi_{\mathbf{3}}(\tau,\bm{z})\,,
 \end{equation}
 where 
 \begin{equation}
     \Phi_{\mathbf{3}}(\tau,\bm{z}) = \Phi^{\bm{\omega}_1}(\tau,\bm{z}) + \Phi^{\bm{\omega}_1-\bm{\alpha_1}}(\tau,\bm{z})+\Phi^{\bm{\omega}_1-\bm{\alpha_1}-\bm{\alpha_2}}(\tau,\bm{z})
 \end{equation}
 and 
 \begin{equation}
  \Phi_{R^-}(\tau,\bm{z}) = \Phi^{-\bm{\alpha}_1}(\tau,\bm{z}) + \Phi^{-\bm{\alpha_2}}(\tau,\bm{z})+\Phi^{-\bm{\alpha_1}-\bm{\alpha_2}}(\tau,\bm{z})\,.
 \end{equation}
Moreover, the corresponding  anomally cancellation conditions hold
 \begin{align}
 \begin{split}
 \frac{1}{2}\chi + 30 \bar{K}_B^2 &= n_{\mathbf{3}}\mathrm{dim}(\mathbf{3}) +[g(\Sigma_\kappa)-1][\mathrm{dim}(\mathbf{adj})-\mathrm{rk}(\mathfrak{a}_2)]\,,\\
 -\bar{K}_B \cdot \frac{b_{\kappa}}{\lambda_\kappa}&= \frac{1}{6}\left(A_{\mathbf{adj}}[1-g(\Sigma_\kappa)]-n_{\mathbf{3}}A_{\mathbf{3}}\right)\,,\\
 \left(\frac{b_\kappa}{\lambda_\kappa}\right)^2 &= \frac{1}{3}\left( n_{\mathbf{3}} C_{\mathbf{3}} + [g(\Sigma_\kappa)-1]C_{\mathbf{adj}}\right)\,,
 \end{split}
 \end{align}
 where $\lambda_{\kappa}=1$, $A_{\mathbf{3}}=1$, $A_{\mathbf{adj}}=6$, $B_{\mathbf{3}}= B_{\mathbf{adj}}=0$,  $C_{\mathbf{3}}=\frac{1}{2}$ and  $C_{\mathbf{adj}}=9$. 

\section{Conclusions and discussion}
\label{sec:6}
The goal of this work has been to provide a purely geometrical understanding of the anomaly cancellation conditions of $\mathcal{N}=(1,0)$ six-dimensional supergravity theories realized by F-theory on elliptically fibered Calabi--Yau threefolds. In the large base volume limit, we showed that the three-point couplings of topological string theory reduce to sums of twisted or deformed Eisenstein series~\cite{Oberdieck:2016nvt} that carry information about the spectrum of such theories. We computed the modular transformations of those three-point couplings in two independent ways. First, we used a subset of monodromy transformations in the stringy K\"ahler moduli space, whose induced action on the complex K\"ahler moduli we indentify with an $\mathrm{SL}(2,\mathbb{Z})$ action. Second, from the modular properties of the relevant quasi-Jacobi forms. Comparing the two descriptions allowed us to extract the corresponding anomaly cancellation conditions of the theories under consideration. Independently, we also derived the anomaly cancellation conditions from a modular anomaly equation for topological four-point functions,  obtained as a consequence of the holomorphic anomaly equations of BCOV theory~\cite{Bershadsky:1993cx}. 

This suggests that the supergravity anomaly cancellation mechanism fits into a more general underlying principle. In our case, there is a reinterpretation of the holomorphic anomaly equations as quantum background independence equations, in which the topological string partition function plays the role of a wave function on the complex moduli space $\mathcal{M}$ of a Calabi--Yau threefold $Y$. 
This wave function arises  from the quantization of the middle cohomology $H^3(Y,\mathbb{R})$, equipped with its natural symplectic form pairing~\cite{Witten:1993ed,Verlinde:2004ck,Aganagic:2006wq,Gunaydin:2006bz,Schwarz:2006br}. 
In this formulation, the holomorphic anomaly equations derive as an infinitesimal consequence of the freedom to choose a given polarisation on $H^3(Y,\mathbb{R})$~\cite{Witten:1993ed}. 
This framework has been crucial in establishing special properties of the topological string partition function, such as the index of meromorphic Jacobi forms that appear as base Fourier coefficients when the Calabi--Yau admits a torus fibration~\cite{Huang:2015ada,Gu:2017ccq,DelZotto:2017mee,Cota:2019cjx,Pioline:2025uov}. 
These index properties have been decisive in arguments for the tower weak gravity conjecture in F-theory compactifications~\cite{Lee:2018spm,Lee:2019tst,Cota:2020zse}, providing an independent perspective on the consistency of this class of theories~\cite{Arkani-Hamed:2006emk,Harlow:2022ich}. It would be worthwhile to deepen our understanding of the wave function approach for topological string theory, both from a physical and a mathematical point of view. 

Let us also note that the base degree zero three- and four-point couplings discussed here encode the Kaluza--Klein tower of states associated with the infinite distance limit in the K\"ahler moduli space reached by shrinking the elliptic fiber~\cite{Lee:2019wij,Rudelius:2023odg}. It is striking that the modular properties of this tower of states determine the anomaly-free consistency of the corresponding effective theories. A similar picture holds in the case of critical strings~\cite{Schellekens:1986yi,Schellekens:1986xh,Lerche:1987qk,Lee:2020gvu}. It would be interesting to gain a more systematic understanding of the connection between emergent string limits and the anomalies of the associated effective theories. 

Interestingly, our approach also captures the anomaly cancellation conditions of the five-dimensional theory realization obtained by compactifying M-theory on an elliptic fibered Calabi--Yau threefold $X$, or equivalently F-theory on $X\times S^1$. In this setup, matching Chern--Simons terms of both effective theory descriptions yields the five-dimensional anomaly cancellation conditions, which agree with their six-dimensional counterpart for the mixed and pure gravitational cases~\cite{Grimm:2013oga}. In our framework, we obtain directly the six-dimensional pure gauge anomaly equations directly from geometrical data, via the quasi-modular behaviour of the topological three- or four-point couplings. In contrast, it was already noted in~\cite{Grimm:2015zea} that the six-dimensional anomaly coefficients and its cancellation mechanism  emerges from the matching of Chern--Simons terms in five dimensions when performing special transformations on Wilson loop parameters, corresponding to elliptic transformations in our language, once again exhibiting the anomalous transformation of quasi-Jacobi forms.  
Nevertheless, unlike~\cite{Grimm:2013oga}, we restricted our attention here to cases in which the zero section of $X$ is holomorphic. We leave for future work the cases with a rational zero section and further related aspects of elliptic transformations. 

Let us emphasize once again that in this work have restricted to smooth elliptic fibered Calabi--Yau threefolds, even though F-theory admits  more types of fibrations, such as elliptic fibrations with $\mathbb{Q}$-factorial terminal singularities, 
or alternatively smooth genus one fibrations in which the geometry does not posses  sections but instead $N$-sections with $N>1$~\cite{Braun:2014oya}. Both types of fibrations can be used to describe discrete gauge symmetries in their correspoding six-dimensional theory realizations, while the $\mathrm{SL}(2,\mathbb{Z})$ modularity associated to the large volume and relative conifold monodromy transformations breaks down to a congruence subgroup  $\Gamma_1(N)\subset\mathrm{SL}(2,\mathbb{Z})$~\cite{Cota:2019cjx,Knapp:2021vkm,Duque:2025kaa,Pioline:2025uov}. It has been observed that,  for genus one fibered Calabi--Yau threefolds  whose six-dimensional effective field theory exhibits no continuous gauge group, the modular properties of the base degree zero topological string partition function yields  constraints analogous to the pure gravitatinal ones discussed here~\cite{Pioline:2025uov}. These constraints were also derived by matching Chern-Simons terms in~\cite{Duque:2025kaa}. See also~\cite{Arras:2016evy,Grassi:2018rva} for gravitational anomaly cancellation conditions for elliptic fibrations with $\mathbb{Q}$-factorial terminal signularities. We contemplate to discuss more general genus one fibrations and its relationship with global anomalies of discrete gauge symmetries~\cite{Dierigl:2022zll} in a companion paper~\cite{FierroCota:2026xxx}.

\appendix 

 \section{Gromov--Witten theory basics}
 \label{App:GW}
 
 In the topological string A-model, Gromov--Witten theory deals with counting holomorphic stable maps $f : C \to X$ 
 from $n$-pointed curves $(C, p_1, \ldots p_n)$ of genus $g$ into a smooth projective variety $X$, such that $f_*[C] = \beta \in H_2(X)$. 
For a nice overview of stable maps see~\cite{Cox:2000vi}. 
Let us denote by $\overline{M}_{g,n}(X,\beta)$ be the moduli space of genus $g$, $n$-pointed stable maps representing the class $\beta \in H_2(X,\mathbb{Z})$.  
 The Gromov--Witten invariants are the rational numbers defined by~\cite{Cox:2000vi}
 \begin{equation}
 N_{g,\beta}^X(\gamma_1, \ldots, \gamma_n) = \int_{[\overline{M}_{g,n}(X,\beta)]^{\mathrm{vir}}} \prod^n_{i=1} \mathrm{ev}_i^*(\gamma_n)\,,
 \end{equation}
 where 
 \begin{equation}
 \mathrm{ev}_i : \overline{M}_{g,n}(X,\beta) \to X
 \end{equation} 
 is the $i$-th evaluation map and $\gamma_i \in H^*(X,\mathbb{Z})$ and $[\cdot]^{\mathrm{vir}}$ denotes the virtual fundamental class. 
 
 For the case that $X$ is a Calabi--Yau threefold, we use $N_{g,\beta}^{X}$ to indicate the case with no insertions. 
 Its relation with Gopakumar--Vafa invariants is given by the multi-covering formula~\cite{Gopakumar:1998ii,Gopakumar:1998jq}
 \begin{equation}
 \sum_{\beta \in H_2(X,\mathbb{Z})} \sum_{g\geq0} N_{g,\beta}^{X} \lambda^{2g-2} Q^\beta = \sum_{\beta \in H_2(X,\mathbb{Z})} \sum_{g\geq 0} n_{g,\beta}^{X} \sum_{m \in \mathbb{N}} \frac{1}{m}\left( \sin\left(\frac{m\lambda}{2} \right) \right)^{2g-2} Q^{m\beta} \,,
 \label{eqn:multicovering}
 \end{equation}
 where $n_{g,\beta}^{X}$ are the Gopakumar--Vafa invariants of genus $g$ and associated curve class $\beta$, 
 which in general are conjectured to be integer-valued. 
 For us, of particular interest are the genus zero invariants related by the formula 
 \begin{equation}
 \sum_{\beta \in H_2(X,\mathbb{Z})}N_{0,\beta}^{X} = \sum_{\beta \in H_2(X,\mathbb{Z})} n_{0,\beta}^{X} \sum_{m \in \mathbb{N}}^\infty \frac{Q^{m\beta}}{m^3} \,. 
 \end{equation}

\section{Three-point coupling transformations from monodromies}
\label{app:GSanomalies}

Here we obtain the relevant $\mathcal{A}_{ABCD}^{\mathrm{GS}}$ from the expressions 
\begin{equation}
   ( C_{ABC}^0 )'= \mathcal{A}_{ABCD}^{\mathrm{GS}}z^D + \tau C_{ABC}
    \end{equation}
    given below.

The \textit{pure abelian gauge} three-point coupling transformation reads
\begin{equation}
(C_{abc}^0)' = -(b_{ab} \cdot b_{cd} +b_{ac} \cdot b_{bd} + b_{ad} \cdot b_{bc})z^d + \tau C_{abc}^0  \,.
\label{eqn:CABC1}
\end{equation} 
Moreover, the \textit{pure non-abelian gauge} three-point coupling
\begin{equation}
    (C^0_{I_\kappa J_\kappa K_\kappa })'=-(\mathcal{C}_{JL,\kappa} \mathcal{C}_{IK,\kappa}+\mathcal{C}_{IL,\kappa} \mathcal{C}_{JK,\kappa} + \mathcal{C}_{KL,\kappa} \mathcal{C}_{IJ,\kappa}) b_\kappa \cdot b_\kappa z^{L_\kappa}+\tau C^0_{I_\kappa' J_\lambda' K_\mu '}\,,
\end{equation}
besides 
\begin{equation}
    (C^0_{I_\kappa J_\kappa K_\lambda})' = (\mathcal{C}_{IJ,\kappa} \mathcal{C}_{KL,\lambda}b_\kappa \cdot b_\lambda)z^{L_\lambda}+\tau C^0_{I_\kappa J_\kappa K_\lambda}\,,
\end{equation}
where $\kappa \neq \lambda$ and
\begin{equation}
(C^0_{I_\kappa J_\lambda K_\mu})' = \tau C^0_{I_\kappa J_\lambda K_\mu} \,,
\end{equation}
where $\kappa \neq \lambda$, $\lambda \neq \mu$ and $\kappa \neq \mu$.

The \textit{pure mixed gauge} three-point couplings read
\begin{align}
    (C^0_{a I_\kappa J_{\lambda}})' & = \tau C^0_{a I_\kappa J_{\lambda}} \\
    (C^0_{a b I_{\kappa}})' & = -\mathcal{C}_{IJ,\kappa} (b_\kappa \cdot b_{ab} )z^{J_{\kappa}} +\tau C_{ab I_\kappa}\,.
\end{align}

\section{Quasi-Jacobi forms}
\label{app:QJac}

In this section we briefly review quasi-Jacobi forms and related automorphic forms. An excellent set of introductory lecture notes is~\cite{OberdieckLectures} and more in-depth  references are~\cite{Oberdieck:2017pqm, Oberdieck:2022khj}. For a physics related discussion and applications see also~\cite{Lee:2019tst, Lee:2020gvu, Lee:2020blx}. 

Let us consider a pair of complex variables $(\tau,z) \in \mathbb{H} \times \mathbb{C}$, where $\mathbb{H}$ is the complex upper-half plane, and let $\Gamma \subset \mathrm{SL}(2,\mathbb{Z})$ be a congruence subgroup of the modular group $\mathrm{SL}(2,\mathbb{Z})$. Moreover, 
we introduce $q : = \exp(2\pi i \tau)$ and $\zeta := \exp(2\pi i z)$. 

A \textit{weak Jacobi form} of weight $k$ and index $m$ is a function $\Phi: \mathbb{H} \times \mathbb{C} \to \mathbb{C}$ such that the following holds
\begin{align}
\label{eqn:mod1}
\Phi\left(\frac{a\tau +b}{c\tau +d},\frac{z}{c\tau +d}\right) &= (c\tau + d)^k e\left[ \frac{c m z^2}{c\tau +d} \right] \Phi(\tau,z) \quad\text{for all} \quad 
\begin{pmatrix} a & b \\ c & d\end{pmatrix} \in \Gamma \,.
\\
\Phi(\tau, z + \lambda \tau + \mu) &= e\left[-m \lambda^2 \tau -2\lambda m z\right] \Phi(\tau,z) \quad \text{for all} \quad \mu, \lambda \in \mathbb{Z}\,.
\end{align}
$\Phi: \mathbb{H} \times \mathbb{C} \to \mathbb{C}$  is holomorphic and admits a Fourier expansion of the form
\begin{align}
\Phi(\tau,z) = \sum_{n\geq0} \sum_{r \in \mathbb{Z}} c(n,r) q^n \zeta^r 
\label{eqn:holExp}
\end{align}
in the region  $\vert q \vert<1$.

For us it is relevant to consider a more general class of automorphic forms that relaxes the notion of holomorphicity. 
Let us introduce the non-holomorphic functions
\begin{equation}
\label{eqn:nuAlpha}
    \nu = \frac{1}{8\pi \mathrm{Im}(\tau)} \,, \quad \alpha = \frac{\mathrm{Im}(z)}{\mathrm{Im}(\tau)} \,.
\end{equation}
An \textit{almost-holomorphic Jacobi form} of weight $k$ and index $m$ is a function $\widehat{\varphi}: \mathbb{H} \times\mathbb{C}\to \mathbb{C}$ such that
\begin{align}
  \widehat{\varphi}\left(\frac{a\tau +b}{c\tau +d},\frac{z}{c\tau +d}\right) &= (c\tau + d)^k e\left[ \frac{c m z^2}{c\tau +d} \right] \widehat{\varphi}(\tau,z) \quad\text{for all} \quad 
\begin{pmatrix} a & b \\ c & d\end{pmatrix} \in \Gamma \,, \\
\widehat{\varphi}(\tau, z + \lambda \tau + \mu) &= e\left[-m \lambda^2 \tau -2\lambda m z\right] \widehat{\varphi}(\tau,z) \quad \text{for all} \quad \mu, \lambda \in \mathbb{Z} \\
\widehat{\varphi }(\tau,z)= & \sum_{i,j \geq0} 
\varphi_{i,j} (\tau,z)\frac{\nu^i}{i!} \frac{\alpha^j}{j!}\,,
\end{align}
where only finitely of the $\varphi_{i,j}$ are non-zero. Each $\varphi_{i,j}$ is a holomorphic function that satisfies and expansion of the form~(\ref{eqn:holExp}) in the region $\vert q \vert <1$.  

A \textit{quasi-Jacobi form} of weight $k$ and index $m$ is a function $\varphi: \mathbb{H}\times \mathbb{C} \to \mathbb{C}$ such that there exists an almost-holomorphic Jacobi form $\widehat{\varphi} = \sum_{i,j} \varphi_{i,j} \frac{\nu^i}{i!} \frac{\alpha^j}{j!}$ of weight $k$ and index $m$ with $\varphi = \varphi_{0,0}$.  
We list next the transformation properties for a quasi-Jacobi form $\varphi$ of weight $k$ and index $m$:
\begin{align}
\begin{split}
\varphi\left(\frac{a\tau +b}{c\tau +d},\frac{z}{c\tau +d}\right) = & (c\tau + d)^k e\left[ \frac{c m z^2}{c\tau +d} \right]  \\
& \times \sum_{i,j \geq0 } \frac{1}{i!}\frac{1}{j!}\left(-\frac{c}{4\pi i(c\tau+d)}\right)^i \left(\frac{c z}{c\tau +d }\right)^j\varphi_{i,j}(\tau,z) 
\end{split}
\end{align}
for all $\begin{pmatrix} a & b \\ c & d \end{pmatrix} \in \Gamma$. Moreover, 
\begin{align}
\varphi(\tau, z + \lambda \tau + \mu) &= e\left[-m \lambda^2 \tau -2\lambda m z\right] \sum_{j \geq 0 } \frac{1}{j!}(-\lambda)^j \varphi_{0,j}(\tau,z)
\end{align}
for all $ \mu, \lambda \in \mathbb{Z}$.

Let us denote by $\mathsf{Jac}_{k,m}(\Gamma)$, $\mathsf{QJac}_{k,m}(\Gamma)$ and $\mathsf{AHJac}_{k,m}(\Gamma)$ the vector spaces of weak Jacobi, quasi-Jacobi and almost-holomorphic Jacobi forms of weight $k$ and index $m$ for the group $\Gamma$, respectively. Moreover, we denote by 
\begin{equation}
\mathsf{QJac}(\Gamma) = \bigoplus_{m\geq0}  \bigoplus_{k\in\mathbb{Z}} \mathsf{QJac}(\Gamma)_{k,m}
\end{equation}
the bi-graded $\mathbb{C}$-algebra of quasi-Jacobi forms and similar for $\mathsf{AHJac}(\Gamma)$ and $\mathsf{Jac}(\Gamma)$.

Notice that the transformation properties for a quasi-Jacobi form $\varphi$ requires the information about the higher order coeffiecients $\varphi_{i,j}$  that define the expansion of an almost-holomorphic Jacobi form.  
In fact, there is a well-defined isomorphism realized by the constant term map 
\begin{equation}
\mathrm{ct} : \mathsf{AHQJac}_{k,m}(\Gamma) \to \mathsf{QJac}_{k,m} (\Gamma)\,, \quad  \widehat{\varphi} = \sum_{i,j \geq0 }  \varphi_{i,j} \frac{ \nu^i}{i!} \frac{ \alpha^j}{j!} \mapsto \varphi_{0,0} \,. 
\label{eqn:ctMap}
\end{equation}
See~\cite{Libgober2009EllipticGR} for a detailed proof. 
\\
\\
\textbf{Example-1:} The \textit{second Eisenstein series} $E_2$ can be defined as follows
\begin{equation}
E_2(\tau) = D_\tau \log( \Delta(\tau)) \,, 
\label{eqn:defEG1}
\end{equation}
 where $D_\tau = \frac{1}{2\pi i}\partial_\tau$ and $\Delta(\tau) = q \prod_{n>0} (1-q^n)^{24}$ is the weight 12 modular cusp form of $\mathrm{SL}(2,\mathbb{Z})$, 
 which transforms as $\Delta(-1/\tau)= \tau^{12} \Delta(\tau)$.
  For convenience, let us introduce the normalized second Eisenstein series $G_2 = -\frac{1}{24}E_2$. 
  From definition~(\ref{eqn:defEG1}) we obtain the following transformation
  \begin{equation}
  G_2(-1/\tau) = \tau^2 G_2(\tau) - \frac{1}{4\pi i} \tau \,. 
  \end{equation}
  From definition~(\ref{eqn:nuAlpha}) we also obtain the following transformation rule
  \begin{equation}
  \nu(-1/\tau) = \tau^2 \nu (\tau) +\frac{\tau }{4\pi i}\,.
  \end{equation}
 We see that 
 \begin{equation}
 \widehat{G}_2 (\tau) = G_2(\tau)  + \nu(\tau) 
 \end{equation}
is in $\mathsf{AHJac}_{2,0}$, 
 while $G_2$ is in $\mathsf{QJac}_{2,0}$.  
 
Note that the space $\mathsf{QJac}_{k,0}$ are simply quasi-modular forms, while their counterparts $\mathsf{AHJac}_{k,0}$ are almost-holomorphic modular forms. 
 See~\cite{Kaneko1995AGJ} for further discussion on quasi-modular forms. 
 \\\\
 \noindent{\textbf{Example-2:}} We consider now the log-derivative of the theta function
 \begin{equation}
A(\tau,z) = D_z \log\vartheta(\tau,z) \,,
\label{eqn:defA}
 \end{equation}
 where 
 \begin{equation}
   \vartheta(\tau,z) = \sum_{\nu \in \mathbb{Z}+\frac{1}{2}}(-1)^{\lfloor 
    \nu \rfloor} q^{\nu^2} \zeta^{\nu} \,.
 \end{equation}
 The theta function has the following transformation properties
 \begin{align} 
 \begin{split}
    \vartheta(\tau+1,z)& =e\left[\frac{1}{8}\right]\vartheta(\tau,z) \,,\quad   \vartheta(-1/\tau,z/\tau) = - i\sqrt{\frac{\tau}{i}} e\left[\frac{z^2}{2\tau}\right] \vartheta(\tau,z) \\
    \vartheta(\tau,z+1) & = -\vartheta(\tau,z)\,, \quad \vartheta(\tau,z+\tau) = -q^{-\frac{1}{2}}\zeta^{-1}\vartheta(\tau,z)\,.
    \end{split}
    \label{eqn:transTheta}
 \end{align}
 Taking definition~(\ref{eqn:defA}) and the transformation properties~(\ref{eqn:transTheta}) we obtain
 \begin{align}
     \begin{split}
         A(-1/\tau,z/\tau) = \tau A(\tau,z) + z  \,, \quad A(\tau,z+\tau)=A(\tau,z) -1 \,. 
     \end{split}
 \end{align}
 From definition~(\ref{eqn:nuAlpha}) we obtain the following transformation rules
 \begin{align}
     \begin{split}
         \alpha(-1/\tau,z/\tau) = \tau \alpha(\tau,z) - z  \,, \quad \alpha(\tau,z+\tau)=\alpha(\tau,z) +1 \,. 
     \end{split}
     \label{eqn:alphaTrans}
 \end{align}
 We see that
 \begin{equation}
     \widehat{A}(\tau,z) = A(\tau,z) + \alpha
 \end{equation}
 is in $\mathsf{AHJac}_{1,0}$, while $A(\tau,z)$ is in $\mathsf{QJac}_{1,0}$. 

 The algebra of quasi-Jacobi forms $\mathsf{QJac}(\Gamma)$ can be embedded in $\mathsf{Jac}(\Gamma) [G_2,A]$~\cite{Oberdieck:2022khj}, where $\mathsf{Jac}(\Gamma)$ is the algebra of weak Jacobi forms. 
 For the case in which $\Gamma = \mathrm{SL}(2,\mathbb{Z})$, 
 we can give a more precise description for $\mathsf{QJac} \equiv \mathsf{QJac}(\mathrm{SL}(2,\mathbb{Z}))$. 
 Let us consider the Weierstra{\ss} elliptic function
 \begin{equation}
     \wp(\tau,z) = \frac{1}{12} + \frac{\zeta}{(1-\zeta)^2} + \sum_{d\geq1} \sum_{k \vert d} (\zeta^k-2+\zeta^{-k}) q^d \,,
     \label{eqn:WP}
 \end{equation}
 as well as its derivative $\wp'(\tau,z) \equiv D_z \wp(\tau,z)$ with respect to the elliptic parameter $z$. 
 Then, 
 \begin{equation}
     \mathcal{R} = \mathbb{C}[\Theta,A,E_2,\wp,\wp',E_4]
     \label{eqn:quasiRing}
  \end{equation}
is a free polynomial ring on its generators and $\mathsf{QJac}$ is equal to the subring of polynomials which define holomorphic functions $\varphi : \mathbb{H}\times \mathbb{C}\to \mathbb{C}$~\cite{vanIttersum:2020lri}. 
Notice in~(\ref{eqn:quasiRing}) the absence of the generator $E_6$ for the ring of  modular forms with modular group $\mathrm{SL}(2,\mathbb{Z})$. 
This is due to the algebraic relation
\begin{equation}
       E_6(\tau) =  18 E_4(\tau)\wp(\tau,z) - 864 \wp^3(\tau,z) + 216 \wp_z^2(\tau,z) \,.
\end{equation}

The algebra $\mathsf{QJac}(\Gamma)$ is closed under the derivative operators $D_\tau:= \frac{1}{2\pi i}\partial_\tau$ and $D_z:= \frac{1}{2\pi i}\partial_z$, where
\begin{align}
\begin{split}
    D_\tau &: \mathsf{QJac}_{k,m}(\Gamma) \to \mathsf{QJac}_{k+2,m}(\Gamma) \,,\\
    D_z & : \mathsf{QJac}_{k,m}(\Gamma) \to \mathsf{QJac}_{k,m}(\Gamma) \,.
    \end{split}
\end{align}
Other relevant operator to us that preserve the $\mathsf{QJac}(\Gamma)$ algebra are the \textit{anomaly operators}
\begin{align}
\begin{split}
    \frac{\partial}{\partial G_2} &: = \mathrm{ct} \circ \frac{\partial}{\partial \nu} \circ \mathrm{ct}^{-1} : \mathsf{QJac}(\Gamma)_{k,m} \to \mathsf{QJac}(\Gamma)_{k-2,m} \,, \\
    \frac{\partial}{\partial A} &: = \mathrm{ct} \circ \frac{\partial }{\partial \alpha} \circ\mathrm{ct}^{-1}: \mathsf{QJac}_{k,m}(\Gamma) \to \mathrm{QJac}_{k-1,m} (\Gamma)\,.
    \end{split}
\end{align}
Note that $\partial_{G_2}$ can be replaced, equivalently, by the more conventional anomaly operator in the physics literature $\partial_{E_2}$, which reads
\begin{equation}
   \frac{ \partial }{\partial E_2}: = -\frac{1}{24} \frac{\partial}{\partial G_2} = \mathrm{ct}\circ\left(-\frac{1}{24}\frac{\partial}{\partial \nu} \right) \circ \mathrm{ct}^{-1} = \mathrm{ct}\circ\left(-\frac{2\pi i}{3}\mathrm{Im}(\tau)^2 \frac{\partial}{\partial \bar{\tau}} \right) \circ \mathrm{ct}^{-1}\,.
   \label{eqn:delE2}
\end{equation}
 \\\\
\noindent\textbf{Example-3:} The $n$-th deformed Eisenstein series $J_n$, introduced in~(\ref{eqn:defEis}), transforms as follows~\cite{2012arXiv1209.5628O}
\begin{align}
\begin{split}
J_n(-1/\tau,z/\tau) & = \sum_{k=0}^n \binom{n}{k} z^{n-k} \tau^k J_k(\tau,z)\,. \\
J_n(\tau,z + \lambda \tau + \mu) & = \sum_{k=0}^n (-1)^{n+k}\binomial{n}{k} \lambda^{n-k} J_k(\tau,z) \text{ for all } \lambda , \mu \in\mathbb{Z} \,.
\label{eqn:nEisTrans} 
\end{split}
\end{align}
The following non-holomorphic function
\begin{equation}
 \widehat{J}_n := \sum_{j = 0}^n \binomial{n}{j} J_{n-j} \alpha^j
 \end{equation}
 is in $\mathsf{AHJac}_{n,0}$, while $J_n$ is in $\mathsf{QJac}_{n,0}$. To prove this, we use the transformation properties of $J_n$ in~(\ref{eqn:nEisTrans}), together with those of $\alpha$ in~(\ref{eqn:alphaTrans}), similarly as in \textbf{Example-2}. 
 In fact, $J_1 = A$.

Another useful algebraic relation for us is the following one
\begin{align}
   D_z A(\tau,z) = &  - \wp(\tau,z) + \frac{1}{12} E_2(\tau)\,.
   \label{eqn:DerGens}
   \end{align}
   We have the following algebraic relations for the following deformed Eisenstein series
   \begin{align}
   \begin{split}
   J_1(\tau,z)  = & A(\tau,z) \\ 
   J_2(\tau,z)  = & A(\tau,z)^2 - \wp(\tau,z)\\
   J_3(\tau,z)  = & A(\tau,z)^3 - 3 A(\tau,z) \wp(\tau,z) - \wp'(\tau,z)\\
   J_4(\tau,z)  = & A(\tau,z)^4 + \frac{E_4(\tau)}{20} - 6 A(\tau,z)^2 \wp(\tau,z) - 3 \wp(\tau,z)^2 \\
   &- 4 A(\tau,z) \wp'(\tau,z)\,.
   \end{split}
\end{align}
Other useful identities for us read below:
\begin{align}
\begin{split}
    D_\tau J_2(\tau,z) =& \frac{1}{6}E_2(\tau)\left[A(\tau,z)^2 - \wp(\tau,z)\right] + \frac{1}{36}E_4(\tau) - 2 A(\tau,z)^2 \wp(\tau,z)  \\
    & - 2 \wp(\tau,z)^2 - 2 A(\tau,z) \wp'(\tau,z)\,,\\
        D_\tau J_4(\tau,z) & = \frac{1}{3} E_2(\tau) \Big[ A(\tau,z)^4 + \frac{E_4(\tau)}{20} - 6 A(\tau,z)^2 \wp(\tau,z) - 3 \wp(\tau,z)^2 \\
   &- 4 A(\tau,z) \wp'(\tau,z) \Big]+ \frac{1}{3}A(\tau,z)^2 E_4(\tau) - 4 A(\tau,z)^4 \wp(\tau,z)\\
   &- \frac{2}{15} E_4(\tau) \wp(\tau,z) - 24 A(\tau,z)^2 \wp(\tau,z)^2 + \frac{12}{5} \wp^3(\tau,z) \\
   &- 8 A(\tau,z)^3 \wp'(\tau,z) 
    - 8 A(\tau,z) \wp(\tau,z) \wp'(\tau,z) - \frac{8}{5} \wp'(\tau,z)^2 \,.
    \end{split}
    \label{eqn:idJ2J4}
\end{align}
\\\\

\newpage
\section{Lie algebras and representation theory relations} 
\label{App:WeylChar}
 

 Let $\mathfrak{g}$ be a complex semisimple Lie algebra. An irreducible finite-dimensional representation $\mathbf{R}$ of $\mathfrak{g}$ can be described as a pair $(\rho, V)$, where $V$ is an irreducible highest weight $\mathfrak{g}$-module 
 and $\rho : \mathfrak{g} \to \mathrm{End}(V)$ is the corresponding Lie algebra homomorphism. There is a decompositon
 \begin{equation}
V = \bigoplus_{\bm{\lambda}} V_{\bm{\lambda}}
 \end{equation}
into weight spaces $V_{\bm{\lambda}}$, such that
\begin{equation}
\rho(\mathcal{T}_I) \ket{\bm{\lambda}} = \lambda_{I} \ket{\bm{\lambda}}
\end{equation}
for all $\ket{\bm{\lambda}}\in V_{\bm{\lambda}}$, where $\{\mathcal{T}_I\}$ denotes the Chevalley basis for the Cartan subalgebra $\mathfrak{h}\subset \mathfrak{g}$. The vectors $\bm{\lambda} = (\lambda_1,\ldots,\lambda_{\mathrm{rk}(\mathfrak{g})})$ are the weights of the module $V$ and the Eigenvalues $\lambda_I$ are the Dynkin labels, which can be obtained from $\lambda_I = \left(\bm{\lambda}, \bm{\alpha}^\vee_I\right)$, where $\left(\cdot\,,\cdot\right)$ denotes the Killing form of $\mathfrak{g}$ and $\bm{\alpha}_I^\vee$ is a simple coroot of $\mathfrak{g}$.   
 
The Weyl character of an irreducible representation $\mathbf{R}$ is a function $\chi_{V}: \mathfrak{h} \to \mathbb{C}$ ~\cite{Fuchs:1997jv} 
\begin{equation}
\chi_{V} (\bm{z}) = \sum_{\bm{\lambda}} m_{\bm{\lambda}} \exp(\left(\bm{\lambda},\bm{z}\right))\,, 
\label{eqn:WeylChar}
\end{equation}
where $\mathfrak{h}$ is the 
where $m_{\bm{\lambda}}\in \mathbb{N}$ is the multiplicity of each weight space $V_{\bm{\lambda}} \subset V$. 
Let $\ket{v_{\bm{\lambda}}}$ be a basis of unitary vectors of $V$. Then, we can also express Weyl characters in the following form
\begin{align}
\begin{split}
    \chi_V(\bm{z}) &= 
\sum_{v_{\bm{\lambda}}}\langle v_{\bm{\lambda}} \mid v_{\bm{\lambda}}\rangle \exp((\bm{\lambda},\bm{z})) = \sum_{v_{\bm{\lambda}}} \langle v_{\bm{\lambda}} \vert \exp[\rho(T(\bm{z}))] \mid v_{\bm{\lambda}} \rangle \\
& = \mathrm{tr}_{V}\exp[\rho(T(\bm{z})]\,.
\end{split}
\label{eqn:altChar}
\end{align}
 Here we introduced the notation $T(\bm{z}) = z^I \mathcal{T}_I$ with $z^I = \left( \bm{\omega}_I, \bm{z}\right)$. 

Using~(\ref{eqn:altChar}), we obtain the following expression
\begin{equation}
 \partial_{z^{I_{1}}} \cdots \partial_{z^{I_n}}  \chi_V(\bm{z}) \Big\vert_{\bm{z}=0} = \mathrm{tr}_{V} \rho(\mathcal{T}_{I_1}) \cdots \rho(\mathcal{T}_{I_n}) \equiv \mathrm{tr}_{\mathbf{R}} \mathcal{T}_{I_1} \cdots \mathcal{T}_{I_n}\,,
 \end{equation}
 where we identified our notation here with that of~\cite{Park:2011ji}. 
 Using the same operation on~(\ref{eqn:WeylChar}) we obtain the following identity
 \begin{equation}
     \mathrm{tr}_{\mathbf{R}} \mathcal{T}_{I_1} \cdots \mathcal{T}_{I_n} = \sum_{\lambda} m_{\bm{\lambda}} \lambda_{I_1} \cdots \lambda_{I_n}\,.
     \label{eqn:traceId}
 \end{equation}

\bibliography{References.bib}
\bibliographystyle{utphys}

\end{document}